\documentclass[showpacs,aps,twocolumn]{revtex4}
\usepackage{bm}
\usepackage{amsmath}
\usepackage{graphicx}
\usepackage{subfigure}
\usepackage[usenames,dvipsnames]{color}
\definecolor{darkblue}{RGB}{0,0,196}
\usepackage[colorlinks=true,linkcolor=darkblue,citecolor=darkblue,urlcolor=darkblue]{hyperref}
\usepackage{setspace}
\usepackage{hyperref}
\usepackage{xcolor}
\usepackage{footmisc}
\usepackage[makeroom]{cancel}
\usepackage{comment}
\usepackage{lineno}
\def\be{\begin{equation}}
\def\ee{\end{equation}}
\def\ba{\begin{eqnarray}}
\def\ea{\end{eqnarray}}
\usepackage{graphicx}
\usepackage{amsmath,bbm}
\usepackage{amssymb,bm}
\usepackage{yfonts}

\begin{document}
\title{Predictions on global properties in O+O collisions at the Large Hadron Collider using a multi-phase transport model}

\author{Debadatta Behera$^{1}$}
\author{Neelkamal Mallick$^{1}$}
\author{Sushanta Tripathy$^{2}$}
\author{Suraj Prasad$^{1}$}
\author{Aditya Nath Mishra$^{3}$}
\author{Raghunath Sahoo$^{1,4}$\footnote{Corresponding author: $Raghunath.Sahoo@cern.ch$}}

\affiliation{$^1$Department of Physics, Indian Institute of Technology Indore, Simrol, Indore 453552, India}
\affiliation{$^2$INFN - sezione di Bologna, via Irnerio 46, 40126 Bologna BO, Italy}
\affiliation{$^3$Wigner Research Center for Physics, H-1121 Budapest, Hungary}
\affiliation{$^{4}$CERN, CH 1211, Geneva 23, Switzerland}

\begin{abstract}
\noindent
Oxygen ($^{16}$O) ions are planned to be injected at the Large Hadron Collider (LHC) in its next runs, and a day of physics run is anticipated for O+O collisions at $\sqrt{s_{\rm{NN}}}$ = 7 TeV. As the system size of O+O collisions has the final state multiplicity overlap with those produced in pp, p+Pb and Pb+Pb collisions, the study of global properties in O+O collisions may provide a deeper insight into the heavy-ion-like behavior observed in small collision systems and its similarities/differences with a larger system like Pb+Pb collisions. In the present work, we report the predictions for global properties in O+O collisions at $\sqrt{s_{\rm{NN}}}$ = 7 TeV using a multi-phase transport model (AMPT). We report the mid-rapidity charged-particle multiplicity, transverse mass, Bjorken energy density, pseudo-rapidity distributions, squared speed of sound, transverse momentum ($p_{\rm T}$) spectra, the kinetic freeze-out parameters, and $p_{\rm T}$-differential particle ratio as a function of collision centrality. Further, we have studied the transverse momentum-dependent elliptic flow of charged particles. The results are shown for Woods-Saxon and harmonic oscillator nuclear density profiles. In addition, we have compared the results with an $\alpha$-clustered structure incorporated inside the oxygen nucleus. Average charged-particle multiplicity and the Bjorken energy density show a significant increase in most central collisions for the harmonic oscillator density profile, while other global properties show less dependence on the density profiles considered in this work. The results from the $\alpha$-clustered structure incorporated inside the oxygen nucleus show similar initial energy density and final charged-particle multiplicity as observed for the harmonic oscillator density profile.

\end{abstract}
\date{\today}
\maketitle 

\section{Introduction}
\label{intro}

In order to understand the properties of the hot and dense medium, often referred to as Quark-Gluon Plasma (QGP), formed in ultra-relativistic heavy-ion collisions at the Large Hadron Collider (LHC) and Relativistic Heavy Ion collider (RHIC), several measurements in different collision systems at different center-of-mass energies are performed. Historically in these collider experiments, heavy-ion collisions such as Pb+Pb and Au+Au collisions are the primary focus in the study of QGP while the small collision systems such as proton-proton (pp) collisions act as a baseline. However, recent results from the LHC experiments~\cite{ALICE:2017jyt,Khachatryan:2016txc} show QGP-like properties in high-multiplicity pp collisions, which raises concerns in the heavy-ion physics community about whether pp collisions can act as a baseline and in addition, if QGP-droplets are produced in the collisions of small systems at the LHC energies. This ambiguity has a serious consequence on the results reported for heavy-ion collisions. Thus, a closer look at the small collision systems is a call of time.

In the upcoming run at the LHC, brief oxygen-oxygen (O+O) collisions are anticipated~\cite{Brewer:2021kiv}, which has a final state multiplicity overlap with pp, p+Pb, and Pb+Pb collisions. Recently, several theoretical studies on O+O collisions have been performed~\cite{Lim:2018huo,Rybczynski:2019adt,Huang:2019tgz,Sievert:2019zjr}. A detailed study on systems formed in O+O collisions may give a deeper insight into understanding the QGP-like properties in small collision systems. As $^{16}$O is doubly magic, it is assumed to be stable against decay and has a very compact structure~\cite{Ropke:2017qck}. Also, $\alpha-$clustered structure~\cite{Li:2020vrg} is proposed to affect oxygen nuclei, where the mean-field effect is not strong enough to break the cluster structure. The $\alpha-$clustering in a nucleus appears when two protons and two neutrons cluster together. A signature of $\alpha-$clustering is seen in recent simulation studies for collisions involving oxygen nuclei~\cite{Li:2020vrg}. However, solid experimental evidences are not yet in place in support of the exotic tetrahedral $\alpha-$clustering structure of $^{16}$O nucleus, which was originally proposed by G. Gamow \cite{Gamow} and then by J.A. Wheeler \cite{Wheeler}. In particular, probing
the $\alpha-$clustering structure in relativistic collisions of light nuclei is being conjectured and studied extensively by Broniowski {\it et al.} \cite{Broniowski-PRL,Broniowski-PRC,Bozek:2014cya}.
In a basic nuclear shell model calculation~\cite{rohlf}, the potential inside the nucleus is assumed to be simple harmonic oscillator potential and then a spin-orbit interaction term is added. From this calculation, the magic numbers are obtained and the nucleus with these magic number of nucleons is expected to be tightly bound and highly stable. As the number of protons and neutrons in oxygen match the magic number individually, the oxygen nucleus is doubly magic and it is expected to be highly compact. Oxygen has also got an isotope ($^{24}$O), which is doubly magic with eight protons and sixteen neutrons but unstable in nature. Readers can refer to
Ref. \cite{Kanungo} for the details of the nuclear structure of $^{24}$O, as an outlook.

To get a deeper insight into the impact of nuclear structure inside the oxygen nucleus, we have incorporated simple harmonic oscillator potential and a more realistic Woods-Saxon potential in the oxygen nucleus using a multi-phase transport model (AMPT)~\cite{AMPT2}. Also, we have compared the results with an $\alpha$-clustered structure incorporated inside the oxygen nucleus. Global observables like charged-particle multiplicities, transverse energy, particle spectra, and pseudorapidity distributions provide insight into the possible formation of QGP in a system. It is proposed that the equation of state of hot hadronic matter can be probed via the correlation of mean transverse momentum and particle multiplicity~\cite{VanHove:1982vk}. Charged-particle multiplicity provides information about the soft processes in the collision, while the mean transverse mass and momentum give insight into the hard processes. In recent results from the LHC~\cite{ALICE:2017jyt}, it has been shown that the QGP-like properties seen in high-multiplicity pp collisions are driven by the final state multiplicity in an event. Thus, it would be interesting to confront initial and final state effects in O+O collisions as it has multiplicity overlap with high-multiplicity pp collisions. The initial energy density can be estimated by using Bjorken hydrodynamic model ~\cite{Bjorken:1982qr}, where one uses the transverse energy or charged-particle multiplicity density in rapidity and mean transverse mass for each collision centrality. To explore the final state effects, a study on particle spectra, kinetic freeze-out parameters, and particle ratios can be studied. In this work, the global properties \cite{Kliemant:2008rh} such as Bjorken energy density, squared speed of sound, particle ratios, and kinetic freeze-out parameters are studied for O+O collisions at $\sqrt{s_{\rm{NN}}}$ = 7 TeV using AMPT for both harmonic oscillator and Woods-Saxon nuclear density profile as well as with an $\alpha$-clustered structure incorporated in oxygen nucleus, in order to explore the effect of the nuclear density profile of the final state observables. 


The paper is organized as follows. We begin with a brief introduction to O+O collisions in Section~\ref{intro}. In Section~\ref{section2}, the detailed event generation methodology with AMPT along with different types of nuclear density profiles are discussed.  In section~\ref{section3} we give a detailed discussion of the obtained
results. Finally, the results are summarized in Section~\ref{section4} 
with possible outlook.

\section{Event Generation and Analysis Methodology}
\label{section2}

In this section, we begin with a brief introduction to the AMPT model and then discuss the charge density profile of oxygen.

\begin{table*} [!hpt]
                \centering 
                \caption{The impact parameter, average number of nucleon participants and average number of nucleon-nucleon binary collisions for different density profiles and in different centrality classes for O+O collisions at $\sqrt{s_{\rm{NN}}}$ = 7 TeV. Centrality selection is done through 
                geometrical slicing, {\it i.e.} from the impact parameter.                 
                 \label{tab:glauber}}
                  \scalebox{0.71}{
                \begin{tabular}{|c |c |c | c| l |c  |c  |c | c  |c  |c  |c | c|}
                \hline
                \textbf{Centrality(\%)} & \multicolumn{4}{c|}{{\textbf{Woods-Saxon}}} & \multicolumn{4}{c|}{{\textbf{harmonic oscillator}}} & \multicolumn{4}{c|}{{\textbf{$\alpha$-cluster}}}\\ 
                \hline
                & $b_{\rm{min}}$(fm) & $b_{\rm{max}}$(fm) & $\langle N_{\rm{part}} \rangle$ $\pm$ rms & $\langle N_{\rm{coll}} \rangle$  $\pm$  rms  & $b_{\rm{min}}$(fm) & $b_{\rm{max}}$(fm) & $\langle N_{\rm{part}} \rangle$  $\pm$  rms  & $\langle N_{\rm{coll}} \rangle$ $\pm$  rms  & $b_{\rm{min}}$(fm) & $b_{\rm{max}}$(fm) & $\langle N_{\rm{part}} \rangle$  $\pm$   rms  & $\langle N_{\rm{coll}} \rangle$  $\pm$  rms \\
                \hline
             	\hline
                0--5             &0      &1.47 &28.00 $\pm$ 2.06  &48.33 $\pm$ 9.43 &0     &1.28  &29.48 $\pm$ 1.61  &63.83 $\pm$ 11.32  &0        &1.30           &29.43 $\pm$ 2.02  &55.12 $\pm$ 8.90\\
                \hline
                5--10           &1.47 &2.08 &25.22 $\pm$ 2.27 &39.96 $\pm$ 7.98 &1.28 &1.83   &27.06 $\pm$ 2.00 &52.64 $\pm$ 9.17  &1.30     &1.86           &26.50 $\pm$ 2.24  &46.56 $\pm$ 7.53\\
                \hline
                10--20         &2.08 &2.94 &21.25 $\pm$ 2.81 &30.17 $\pm$ 7.34 &1.83 &2.58   &23.20 $\pm$ 2.68 &39.31 $\pm$ 8.57  &1.86    &2.63            &22.25 $\pm$ 2.57  &36.19 $\pm$ 7.09\\
                \hline
                20--30         &2.94 &3.59 &16.46 $\pm$ 3.01 &20.24 $\pm$ 5.96 &2.58 &3.17   &18.30 $\pm$ 2.94 &26.21 $\pm$ 6.95  &2.63    &3.22            &17.51 $\pm$ 2.49  &25.85 $\pm$ 5.74\\
                \hline
                30--40         &3.59 &4.14 &12.55 $\pm$ 3.20 &13.63 $\pm$ 5.07 &3.17 &3.66   &14.12 $\pm$ 3.17 &17.55 $\pm$ 6.00  &3.22     &3.72             &13.71 $\pm$ 2.42  &18.31 $\pm$ 5.17\\
                \hline
                40--50         &4.14 &4.63 &9.39   $\pm$ 3.15 &9.13   $\pm$ 4.22 &3.66 &4.10   &10.46 $\pm$ 3.21 &11.33 $\pm$ 4.98  &3.72     &4.15             &10.58 $\pm$ 2.43  &12.80 $\pm$ 4.69\\
                \hline
                50--60         &4.63 &5.09 &7.01   $\pm$ 2.86 &6.17   $\pm$ 3.40 &4.10 &4.51   &7.76  $\pm$ 2.99 &7.43   $\pm$ 3.98  &4.15     & 4.56             & 8.05 $\pm$  2.46 &8.74 $\pm$ 4.21\\
                \hline
                60--70         &5.09 &5.54 &5.32   $\pm$ 2.45 &4.27   $\pm$ 2.65 &4.51 &4.89   &5.76  $\pm$ 2.57 &4.96   $\pm$ 3.07  &4.56     &4.94                &6.11 $\pm$ 2.30 &5.98 $\pm$ 3.51\\
                \hline
                70--100       &5.54 &13.46    &3.33   $\pm$ 1.68 &2.28   $\pm$ 1.68 &4.89 &9.86   &3.47  $\pm$ 1.76 &2.47   $\pm$ 1.86  &4.94    &8.79            &3.74 $\pm$ 1.81  &2.97 $\pm$ 2.35\\
                \hline
                \end{tabular}     
                }          
\end{table*}

\begin{table}[!hpt]
                \centering
                \caption{ Centrality dependent average charged-particle multiplicity density for different nuclear profiles in O+O collisions at $\sqrt{s_{\rm{NN}}}$ = 7 TeV in the range  $|\eta|<0.5$ }
                \scalebox{0.89}{
                \begin{tabular}{|c |c |c | c|}

                \hline
               Centrality & Woods-Saxon & harmonic oscillator  & $\alpha-$cluster\\
                \hline
                & $\langle{dN_{ch}}/d{\eta}\rangle$  & $\langle{dN_{ch}}/d{\eta}\rangle$ & $\langle{dN_{ch}}/d{\eta}\rangle$\\
                \hline
             	\hline
                    0--5             &161.07 $\pm$ 0.15    &192.60 $\pm$  0.20      &187.54 $\pm$ 0.14 \\
                    \hline
                    5--10           &139.38 $\pm$ 0.13       &167.37 $\pm$  0.18       &163.86 $\pm$  0.11 \\
                    \hline
                    10--20           & 112.20 $\pm$ 0.13       &134.72 $\pm$ 0.18      &132.67 $\pm$  0.10 \\
                    \hline
                    20--30           &82.84 $\pm$ 0.11       & 99.34 $\pm$ 0.12      &100.21 $\pm$ 0.09 \\
                    \hline
                    30--40           &61.33 $\pm$  0.10     & 73.40 $\pm$ 0.11     & 74.60 $\pm$ 0.08 \\
                    \hline
                    40--50           &45.10 $\pm$ 0.09     &53.23 $\pm$ 0.09        & 54.72 $\pm$ 0.07 \\
                    \hline
                    50--60            &32.81 $\pm$ 0.07     &38.31 $\pm$ 0.08          & 39.56 $\pm$ 0.06 \\
                    \hline
                    60--70            &23.44 $\pm$ 0.06    & 27.38 $\pm$ 0.07          &  27.65  $\pm$ 0.05 \\
                    \hline
                    70--100          &9.85 $\pm$ 0.03       &10.65 $\pm$ 0.04        &10.04 $\pm$ 0.03 \\
                    \hline
                \end{tabular} 
                }              
\end{table}

\subsection{A Multi-Phase Transport (AMPT) Model}
AMPT model primarily consists of four components~\cite{AMPT2}: 1. initialization of the collisions using HIJING~\cite{ampthijing}, where the differential cross-section of the produced mini-jets in pp collisions is calculated and converted to heavy-ion collisions with the inbuilt Glauber model, 2. the produced partons are propagated into parton transport part via Zhang’s Parton Cascade (ZPC) model~\cite{amptzpc}, 3. hadronization mechanism: in AMPT string melting version, the transported partons are hadronized using spatial coalescence mechanism~\cite{Lin:2001zk,He:2017tla}, 4. hadron transport: the produced hadrons undergo a final evolution in relativistic transport mechanism~\cite{amptart1, amptart2} via baryon-baryon, meson-baryon, and meson-meson interactions. Although there is also a default version available in AMPT, we have used the string melting mode of AMPT (AMPT version 2.26t9b) in the current work due to the fact that the particle flow and spectra at the intermediate-$p_{\rm T}$ regions are well explained by a quark coalescence mechanism available in the string melting version for hadronization~\cite{Greco:2003mm,ampthadron2,ampthadron3}. We have used similar AMPT settings in the current work as reported in Ref.~\cite{Tripathy:2018bib} unless specified explicitly. As there is no experimental data available for O+O collisions, we have compared the charged-particle $p_{\rm T}$-spectra from minimum bias p+Pb collisions at $\sqrt{s_{\rm{NN}}}$ = 5.02 TeV in midrapidity with the predictions from AMPT by fixing the partonic scattering cross
section, $\sigma_{\rm{gg}}$ = 3 mb \cite{Lim:2018huo}, which can be seen in the Appendix section A of the paper.

\begin{figure}[ht!]
\includegraphics[scale=0.4]{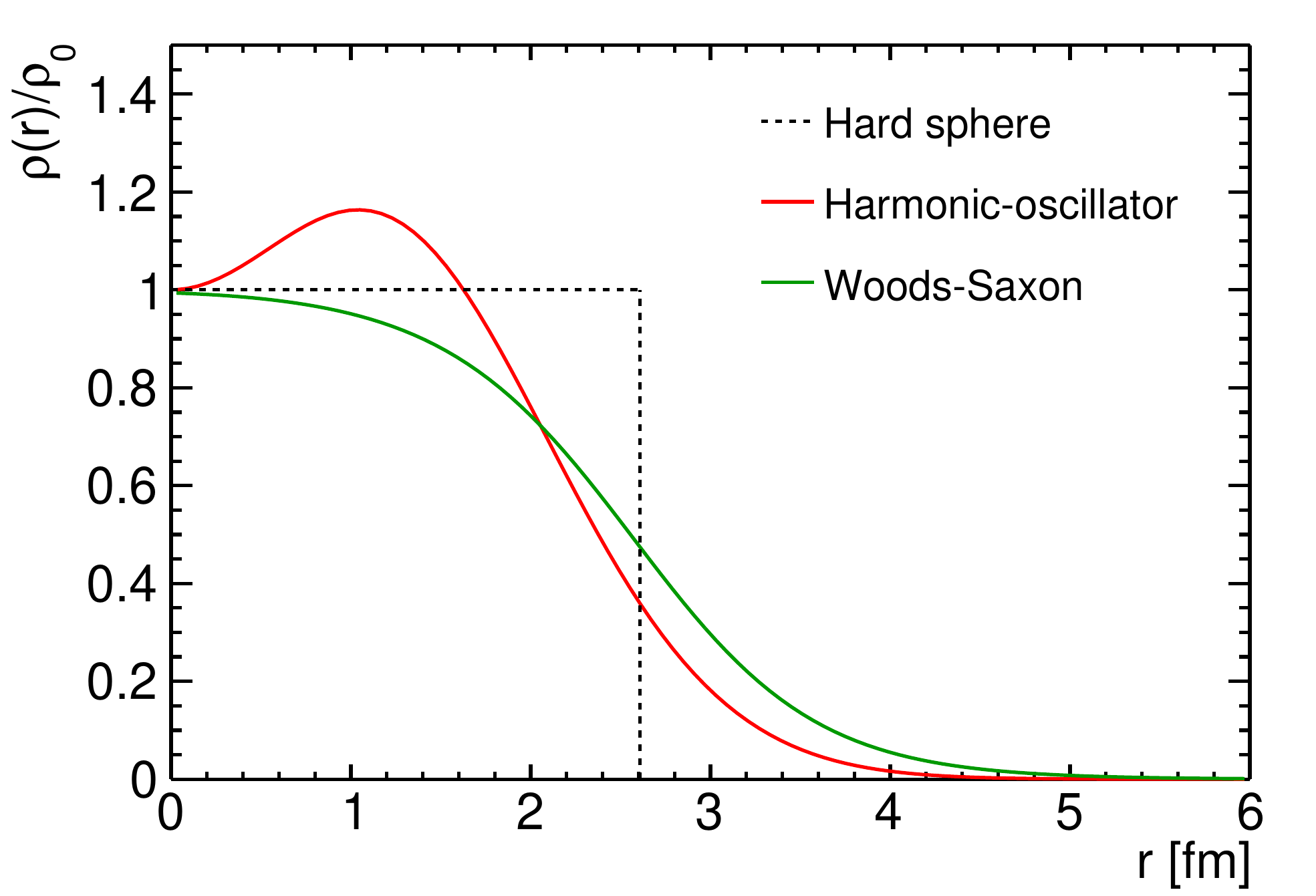}
\caption[]{(Color online) Nuclear density profile for oxygen
nucleus. Shown are the hard sphere, harmonic oscillator and Woods-Saxon density profiles.}
\label{fig1}
\end{figure}

The collision centrality, the number of participants ($N_{\rm{part}}$), and the number of binary collisions ($N_{\rm{coll}}$) in heavy-ion collisions cannot be determined directly from experiments, instead it is obtained via the impact parameter determination in the Glauber model~\cite{Miller:2007ri,Glauber:1970jm, Loizides:2016djv}. The Glauber model considers a nucleus-nucleus collision as a superposition of several independent nucleon-nucleon collisions and the estimations are dependent on the nuclear density profile. In this work, we have modified the inbuilt Glauber model in AMPT to incorporate a harmonic oscillator density profile and $\alpha$-clustered structure in the oxygen nucleus. They are compared with the results with that of the Woods-Saxon nuclear density profile. The values of impact parameters for different centrality classes in O+O collisions are obtained by using the publicly available MC Glauber code (TGlauberMC-3.2)~\cite{Loizides:2017ack,MCglauber,Alver:2008aq,Loizides:2014vua}. The values of the impact parameter, the average number of nucleon participants, and  average number of binary nucleon-nucleon collisions for different density profiles and in different centrality classes are shown in Table~\ref{tab:glauber}.

\subsection{Woods-Saxon density profile}
In high-energy heavy-ion collisions, the standard method employed for nuclear density profile is in terms of three-parameter Fermi (3pF) distribution, which is often referred as Woods-Saxon distribution. The Woods-Saxon distribution is given by,
\begin{equation}
\rho (r) = \frac{\rho_{0}(1+w(\frac{r}{r_0})^2)}{1+{ \rm exp}(\frac{r-r_0}{a})}.
\end{equation}
Here, $r$ is the radial distance from the center of the nucleus, $r_0$ is the mean radius of the nucleus, $a$ is the skin depth/diffusivity of the nucleus and $w$ is the deformation parameter. For the oxygen nucleus, $r_0$ is 2.608 fm, $a$ is 0.513 fm, $w$ is -0.051~\cite{Loizides:2014vua}. $\rho_0$ is the nuclear density constant at $r$ = 0, which is obtained by the overall normalization condition,
\begin{equation}
\int\rho(r) d^3r = 4\pi\int \rho(r)r^2dr = Ze.
\label{rho-norm}
\end{equation}
Here, $Z$ is the atomic number of the nucleus $i.e.,$ 8 for oxygen nucleus. For a hard sphere configuration in $r < r_0$, $\rho(r) = \rho_{0}$, and $\rho_{0} = 3Ze/(4\pi R^3)$.

\subsection{Harmonic oscillator density profile}
The harmonic oscillator charge density distribution is given by,
\begin{equation}
\rho (r) = \rho_{0}\bigg[1 + \alpha\bigg(\frac{r}{a}\bigg)^{2}\bigg]{\rm exp}\bigg(\frac{-r^2}{a^2}\bigg).
\end{equation}
Here, $a$ and $\alpha$ are parameters which are taken as 1.544 fm and 1.833 for oxygen~\cite{Loizides:2014vua}, respectively. Similar to the
Woods-Saxon case, $\rho_0$ is the nuclear density constant at $r$ = 0, which should satisfy the normalization condition mentioned in Eq.~\ref{rho-norm}.

Figure~\ref{fig1} shows the comparison of different normalized nuclear density profiles of oxygen, namely the hard-sphere, Woods-Saxon, and the harmonic oscillator. A significant rise of the normalized density profile at small $r$ is due to the harmonic oscillator potential seen with respect to the Woods-Saxon density profile. As the tetrahedral  $\alpha$-cluster structure of $^{16}$O nucleus is implemented numerically in AMPT, a nuclear density profile is difficult to obtain analytically. However, the probability of the radial position of the nucleons distributed inside the nucleus has been shown in Appendix section B and compared the same for harmonic oscillator potential and the Woods-Saxon density profile. A brief discussion on the details of tetrahedral $\alpha$-cluster structure is mentioned below. 


\subsection{$\alpha$-clustered structure in O$^{16}$}
\begin{figure}[ht!]
\includegraphics[scale=0.4]{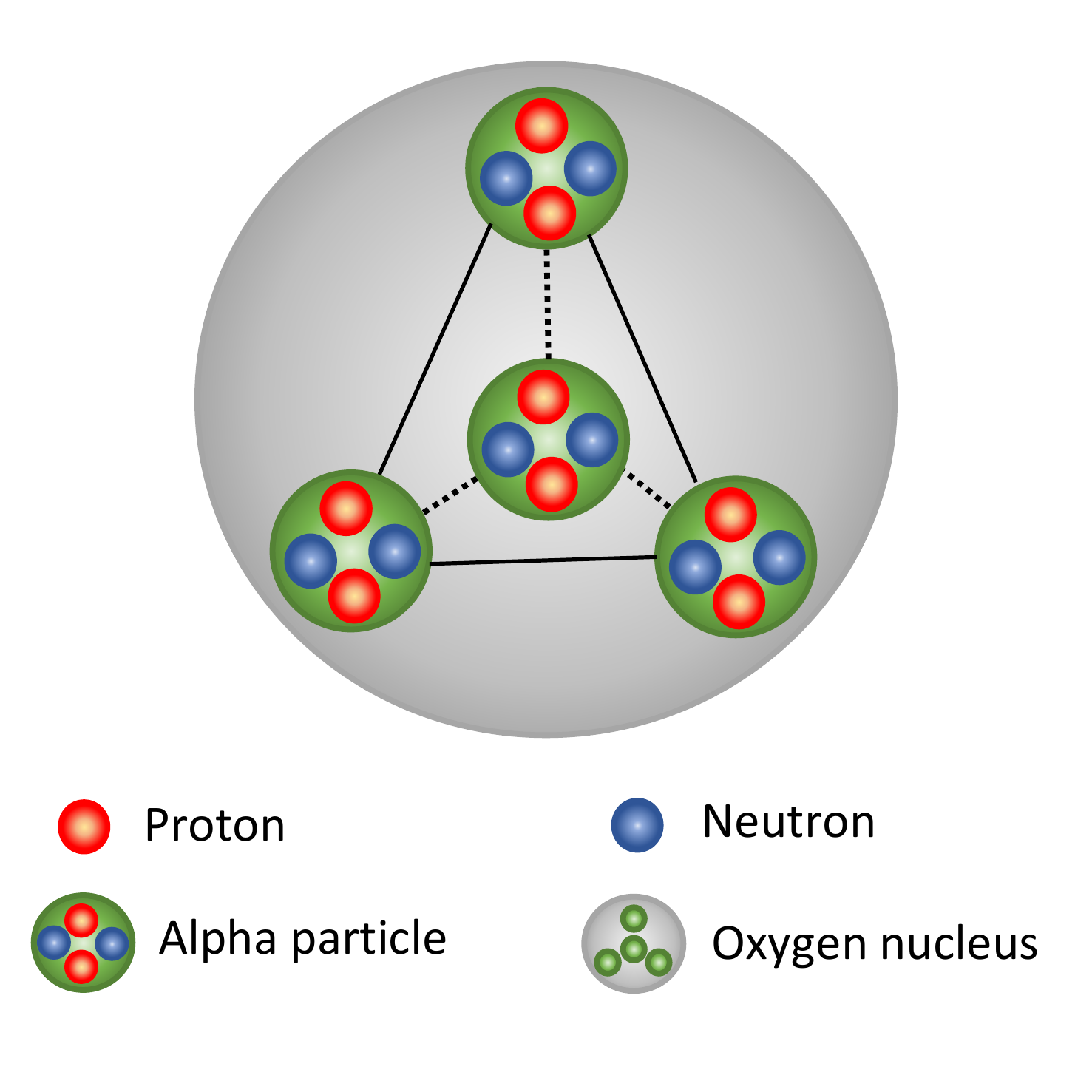}
\caption[]{(Color online) Depiction of $\alpha$-clustered structure in oxygen nucleus.}
\label{fig1.1}
\end{figure}
Clustering plays a crucial role in studying nuclear structure. Protons and neutrons in many-body nuclear systems tend to form clusters in order to reduce overall energy or boost system stability.
Two neutrons and two protons in a nucleus can cluster together to form $\alpha$ particles. The $\alpha$-clustering is observed in several light nuclei such as $^{8}$Be and $^{12}$C~\cite{Tohsaki:2001an}. In these types of nuclei, the mean-field is not strong enough to break the cluster effect. The intrinsic state of the $^{12}$C nucleus is a triangular structure made up of 3 $\alpha$-particles~\cite{Broniowski:2013dia}. Several studies~\cite{Tohsaki:2001an, Barbui:2018sqy} have indicated that the structure in $^{16}$O corresponds to a state analogous to the $^{12}$C state with $\alpha$-clustering. The experimental observation assumes that the $^{16}$O has an $\alpha$-like cluster at the corners of a tetrahedron (Fig.~\ref{fig1.1}). A signature of $\alpha-$clustering is seen in recent simulation studies for collisions involving oxygen nuclei~\cite{Li:2020vrg}. These studies have inspired us to implement the $\alpha-$clustered structure inside the oxygen nucleus using the AMPT model. Nucleons inside an $\alpha-$cluster are distributed following the Woods-Saxon distribution for $^4$He nucleus with a rms radius 1.676 fm. Such randomized $\alpha-$clusters are placed on the vertices of a regular tetrahedron with a side length 3.42 fm. The rms radius of such an arrangement gives the rms radius for $^{16}$O to be 2.699 fm~\cite{Li:2020vrg}. The arrangement of nucleons is randomized event by event by rotating the system along x-y-z directions for both projectile and target nuclei following the tetrahedral structure.

 
\section{Results and Discussions}
\label{section3}
Here, we discuss the predictions for different global properties such as Bjorken energy density, speed of sound, freeze-out parameters as a function of centrality class in different subsections. We have also studied the transverse momentum-dependent elliptic flow of charged particles. Here onwards for simplicity, we refer $\pi^{+}+\pi^{-}$, K$^+$+K$^{-}$ and p+$\bar{\rm p}$ as pions, kaons, and protons, respectively.

\subsection{Transverse energy and Bjorken Energy Density}

\begin{figure}[ht]
\includegraphics[scale=0.40]{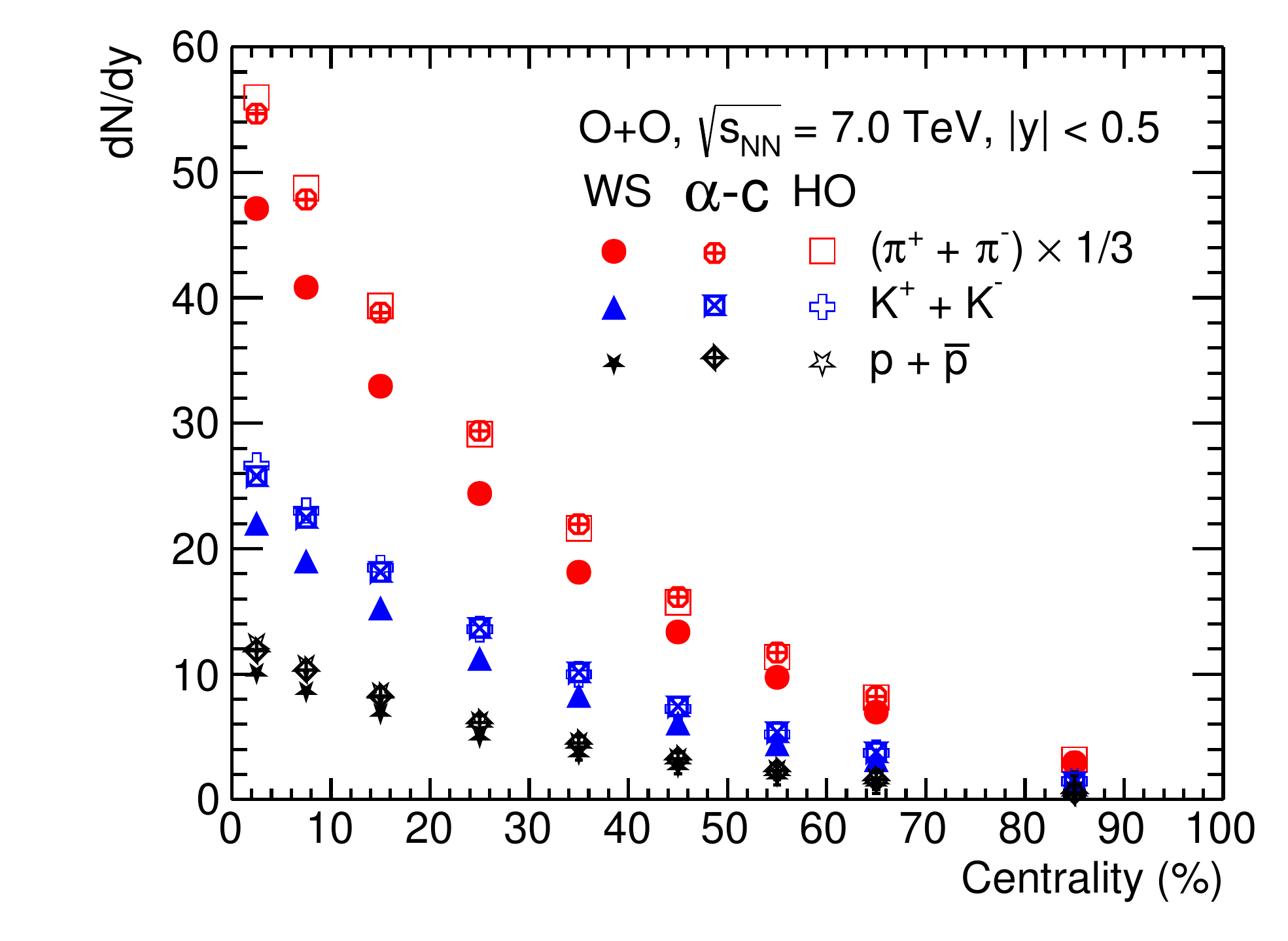}
\caption[]{(Color online) Integrated yield of pions, kaons, and protons at mid-rapidity as a function centrality classes for pions, kaons, and protons in O+O collisions at $\sqrt{s_{\rm{NN}}}$ = 7 TeV. The solid markers represent the Woods-Saxon (WS) density profile; open markers represent the harmonic oscillator (HO) density profile, and the markers with a cross represent the results from $\alpha$-clustered structure.}
\label{fig2}
\end{figure}

\begin{figure}[ht]
\includegraphics[scale=0.40]{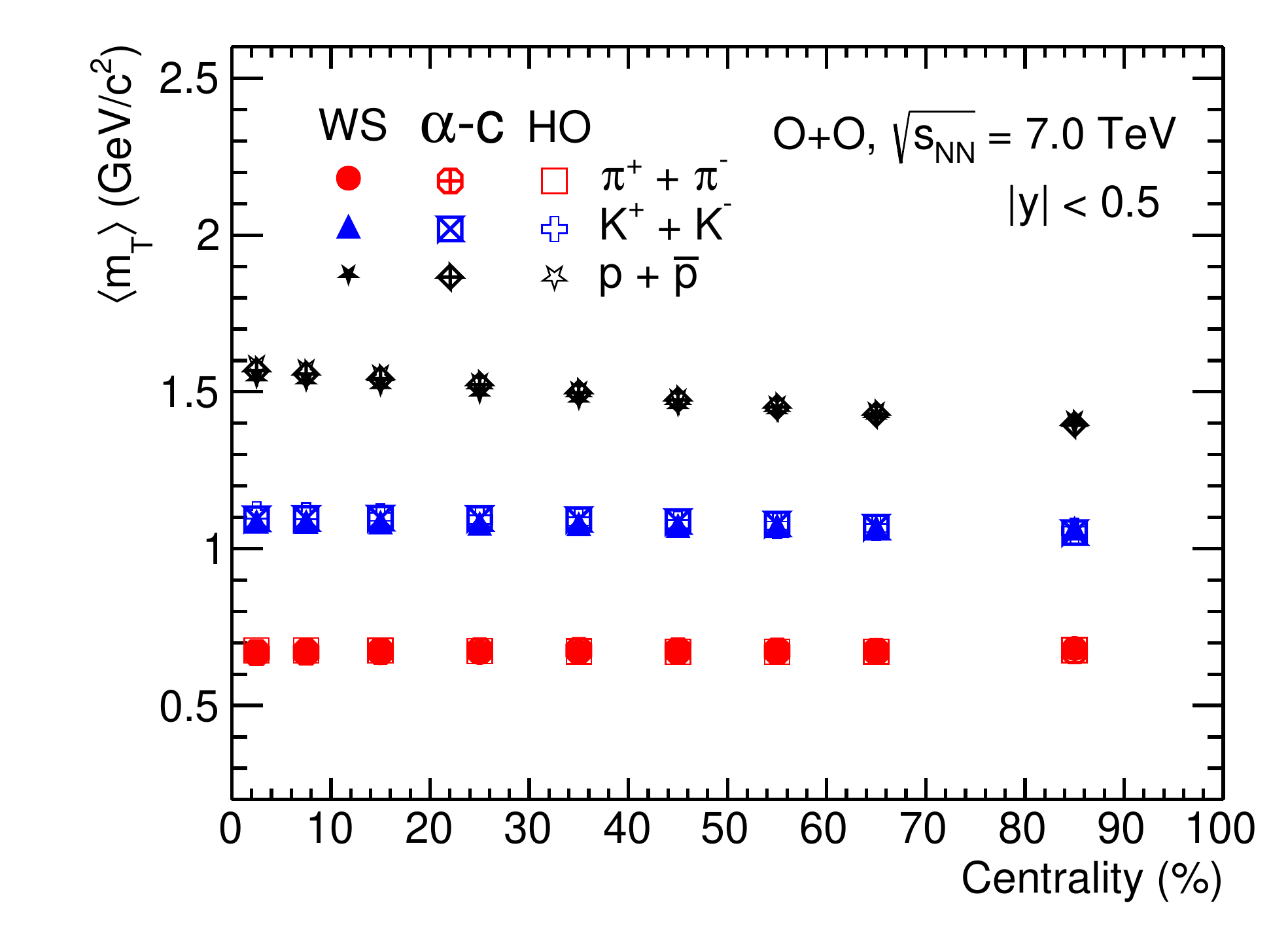}
\caption[]{(Color online) Mean transverse mass as a function centrality classes for pions, kaons and protons in O+O collisions at $\sqrt{s_{\rm{NN}}}$ = 7 TeV.  The solid markers represent the Woods-Saxon (WS) density profile and open markers represent harmonic oscillator (HO) density profile, and the markers with a cross represent the results from $\alpha$-clustered structure.}
\label{fig3}
\end{figure}

\begin{figure}[ht]
\includegraphics[scale=0.40]{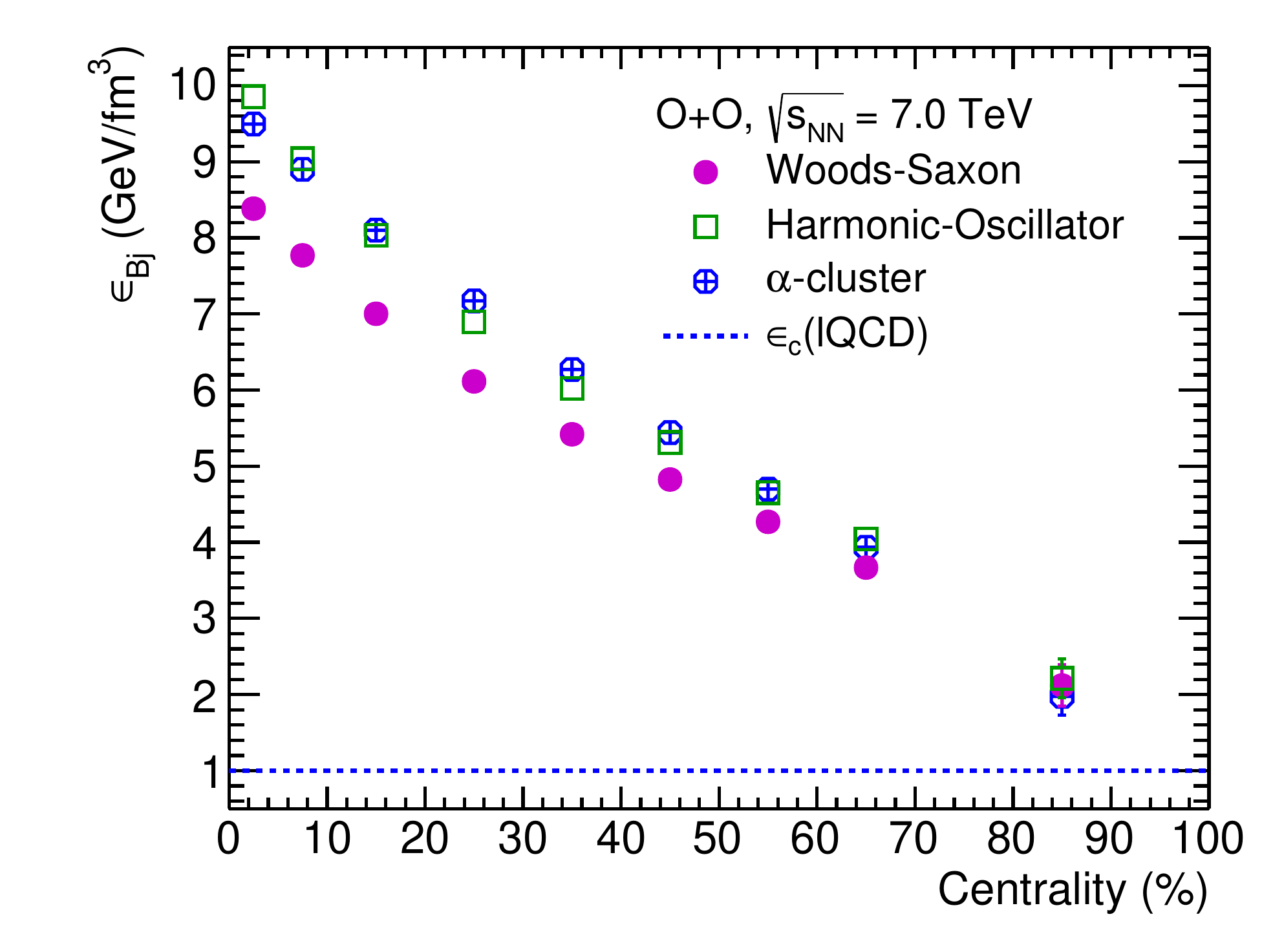}
\caption[]{(Color online) Bjorken energy density as a function centrality classes for pions, kaons and protons in O+O collisions at $\sqrt{s_{\rm{NN}}}$ = 7 TeV. The solid (open) markers represent the Woods-Saxon (harmonic oscillator) density profile, and the markers with a cross represent the results from $\alpha$-clustered structure.}
\label{fig4}
\end{figure}

In the study of QGP properties in heavy-ion collisions, one of the key variables is the initial energy density produced in such collisions. The initial density can be estimated via the Bjorken boost-invariant hydrodynamics model~\cite{Bjorken:1982qr}, where  transverse energy density ($E_{\rm T}$) in mid-rapidity gives the quantitative estimation of the initial energy density produced in the interaction. The Bjorken energy density ($\epsilon_{\rm Bj}$) with the assumption of boost invariance is given as,
\begin{equation}
\epsilon_{\rm{Bj}}  = \frac{1}{\tau S_{\rm T}}\frac{dE_{\rm T}}{dy},
\label{eq-bj}
\end{equation}
where, $S_{\rm T}$ is the transverse overlap area of the two colliding nuclei and $dE_{\rm T}/dy$ is the transverse energy density at midrapidity at a formation time $\tau$. As Eq.~\ref{eq-bj} diverges at $\tau\rightarrow0$, a finite formation time ($\tau$) = 1 fm/c is assumed for the calculation of Bjorken energy density in this work. $E_{\rm T}$ is the total transverse energy produced in an event and $S_{\rm T} = \pi R^2$ is the total transverse overlap area of the colliding nuclei. Replacing, $R = R_{0} A^{1/3}$ and $A = N_{\rm{part}}/2$,
\begin{eqnarray}
S_{\rm T} = \pi R_{0}^2 \left( \frac{N_{\rm{part}}}{2}\right)^{2/3}
\label{transarea}
\end{eqnarray}
As most of the transverse energy is carried by the pions, kaons and protons due to their abundance, the total transverse energy ($E_{\rm T}$) can be approximated as~\cite{ALICE:2016igk,Sahoo:2014aca,STAR:2008med},
\begin{equation}
\begin{gathered}
\frac{dE_{\rm T}}{dy} \approx  \frac{3}{2}\times\left(\langle m_{\rm T} \rangle \frac{dN}{dy}\right)_{\pi^{\pm}} + 2\times\left(\langle m_{\rm T} \rangle\frac{dN}{dy}\right)_{K^{\pm},p,\bar{p}}.
\label{transenergy}
\end{gathered}
\end{equation}
The multiplicative factor in each term accounts for corresponding neutral particles. $m_{\rm T} = \sqrt{p_{\rm T}^2 + m^2}$, is the transverse mass and $dN/dy$ is the integrated yield for $\pi^{\pm}$, $K^{\pm}$ and $p+\bar{p}$ in mid-rapidity region {\em i.e.} $|y| < 0.5$. Now, Eq.~\ref{eq-bj} can be written as,
\begin{equation}
\begin{gathered}
\epsilon_{\rm{Bj}} \approx \frac{1}{\tau\pi R_{0}^2 \left( \frac{N_{\rm{part}}}{2}\right)^{2/3}}\bigg[\frac{3}{2}\times\left(\langle m_{\rm T} \rangle \frac{dN}{dy}\right)_{\pi^{\pm}}\\
 + 2\times\left(\langle m_{\rm T} \rangle\frac{dN}{dy}\right)_{K^{\pm},p,\bar{p}} \bigg]
\label{eq-Bj-final}
\end{gathered}
\end{equation}

Figures~\ref{fig2} and~\ref{fig3} show the integrated yields and mean transverse momenta for pions, kaons and protons as a function of centrality classes for O+O collisions at $\sqrt{s_{\rm{NN}}}$ = 7 TeV,  respectively. The solid markers represent the Woods-Saxon density profile, the open markers represent the results for the harmonic oscillator density profile, and the markers with a cross represent the results from $\alpha$-clustered structure. As expected, pions' integrated yield is higher than kaons and protons as they are the most abundant among the identified particles and are understood from a thermalized Boltzmannian production of secondaries in nuclear collisions. The comparison of Woods-Saxon and harmonic oscillator density profiles show that the soft particle production is higher for the harmonic oscillator density profile in central collisions. The results from the $\alpha$-clustered structure show similar behavior as seen for the harmonic oscillator density profile. However, the mean transverse mass remains nearly the same, indicating a similar spectral shape for all cases. For peripheral collisions, the differences among the density profiles and $\alpha$-clustered structure diminish.

With the input of integrated yields and mean transverse mass for pions, kaons, and protons, the Bjorken energy density is obtained as a function centrality classes using Eq.~\ref{eq-Bj-final}, which is shown in Fig.~\ref{fig4}. The solid markers represent the Woods-Saxon density profile, the open markers represent the harmonic oscillator density profile and the markers with a cross represent the results for the $\alpha$-clustered structure. The Bjorken energy density is found to be higher for most central collisions and linearly decreases while going from central to peripheral collisions. As evident in Eq.~\ref{eq-Bj-final}, Bjorken energy density strongly depends on the integrated yield, and the difference in integrated yield for different nuclear density profiles is reflected in the Bjorken energy density. The oxygen nucleus with harmonic oscillator density profile show about 15\% higher energy density compared to the Woods-Saxon density profile. Thus, for central O+O collisions, the density profile plays a crucial role in studying the initial-state (Bjorken energy density) and final-state (integrated yields) effects. However, going towards the higher impact parameter, $i.e.,$ for peripheral collisions the difference is negligible. As observed for integrated yield, $\alpha$-clustered structure shows similar behavior as seen for harmonic oscillator density profile. The nuclear density profile has a clear effect on the initial energy density, which in fact controls the particle production and the 
subsequent space-time evolution of the fireball, and hence the equation of state (EoS). The values of the initial energy densities 
for all collision centralities are observed to be higher than the lattice QCD estimated requirement of 1 GeV/fm$^3$ energy density
for a deconfinement transition \cite{Karsch:2001vs}. This hints at observing the signals of QGP in oxygen-oxygen collisions at the LHC energies.
We shall further strengthen these arguments in the following sections while discussing other global observables
in heavy-ion collisions.

\subsection{Pseudorapidity distributions and squared speed of sound}

\begin{figure}[ht]
\includegraphics[scale=0.40]{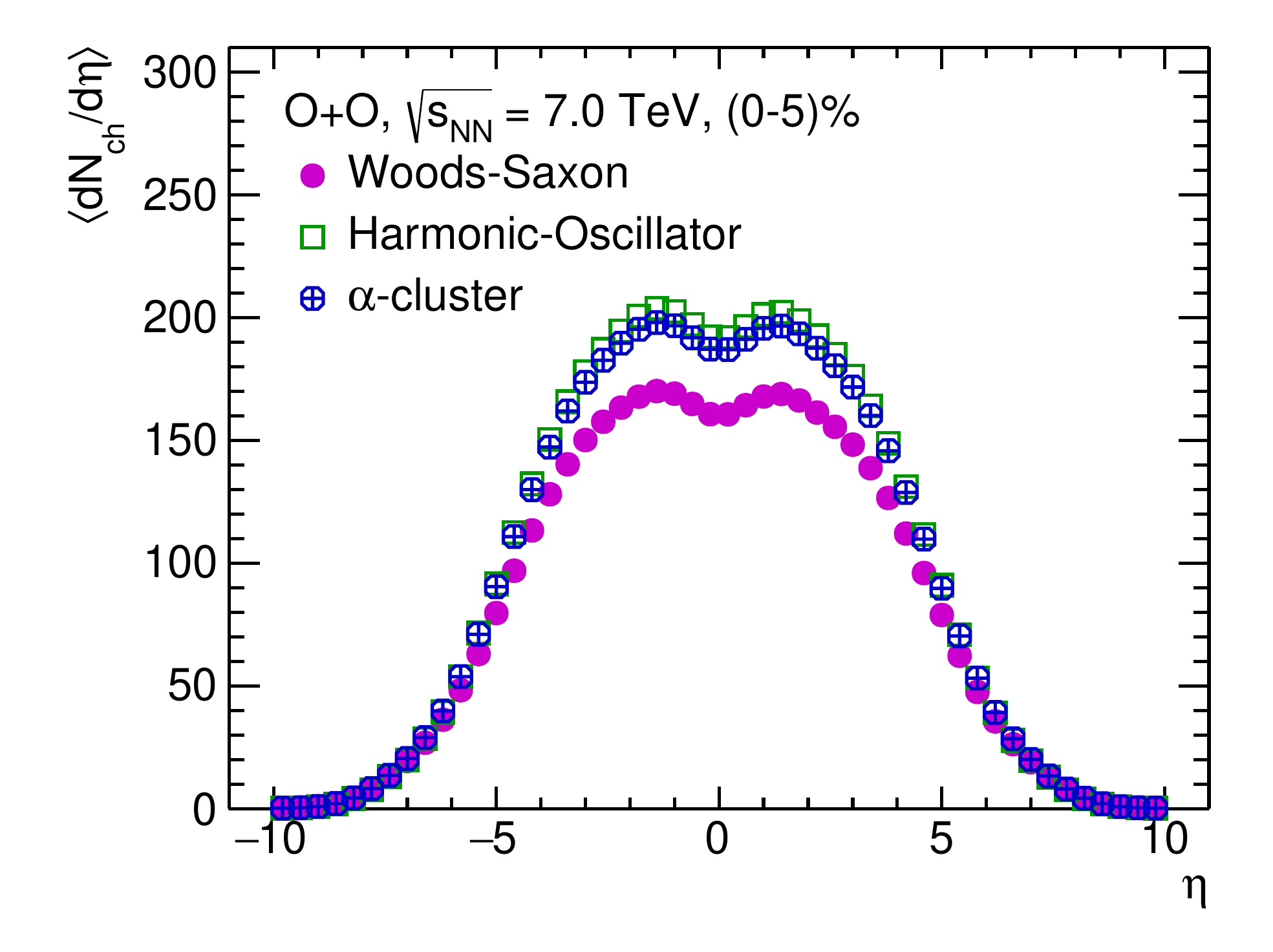}
\includegraphics[scale=0.40]{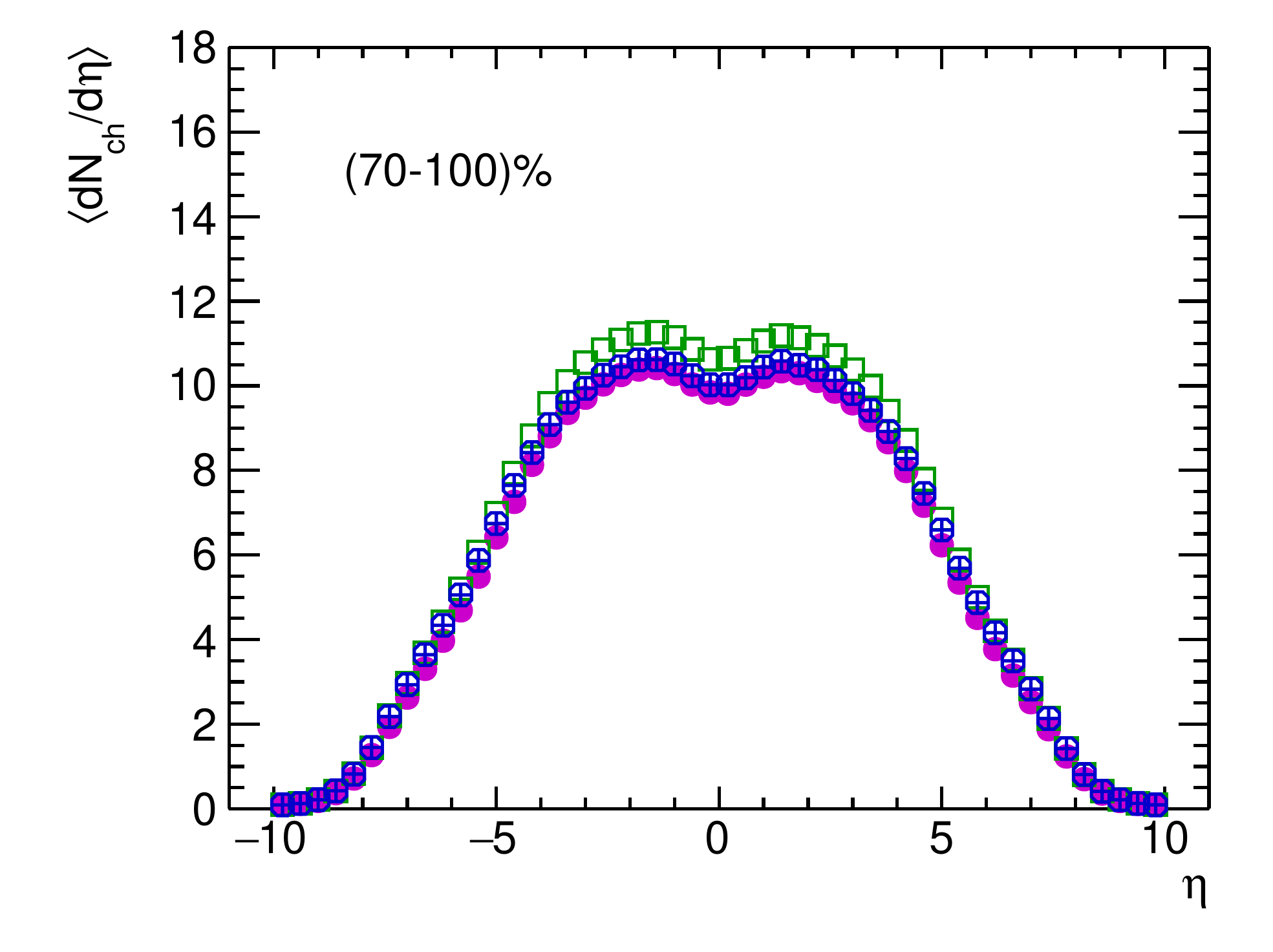}
\caption[]{(Color online) Pseudorapidity distributions of charged particles in O+O collisions at $\sqrt{s_{\rm{NN}}}$ = 7 TeV for (0-5)\% (top) and (70-100)\% (bottom) centrality classes.  The solid (open) markers represent the Woods-Saxon (harmonic oscillator) density profile and the markers with a cross represent $\alpha$-clustered structure.}
\label{fig5}
\end{figure}
Figure~\ref{fig5} shows pseudorapidity distributions of charged particles in O+O collisions at $\sqrt{s_{\rm{NN}}}$ = 7 TeV for (0-5)\% and (70-100)\% centrality classes for Woods-Saxon density profiles, harmonic oscillator density profiles and $\alpha$-clustered structure. Here, the differences in the charged-particle multiplicity in the mid-pseudorapidity due to modification of nuclear density profiles are clearly visible for central collisions while the difference is relatively smaller for peripheral collisions. However, at forward-pseudorapidity, almost no dependence of charged-particle density on the density profiles and $\alpha$-clustered structure is seen.

\begin{figure}[ht]
\includegraphics[scale=0.40]{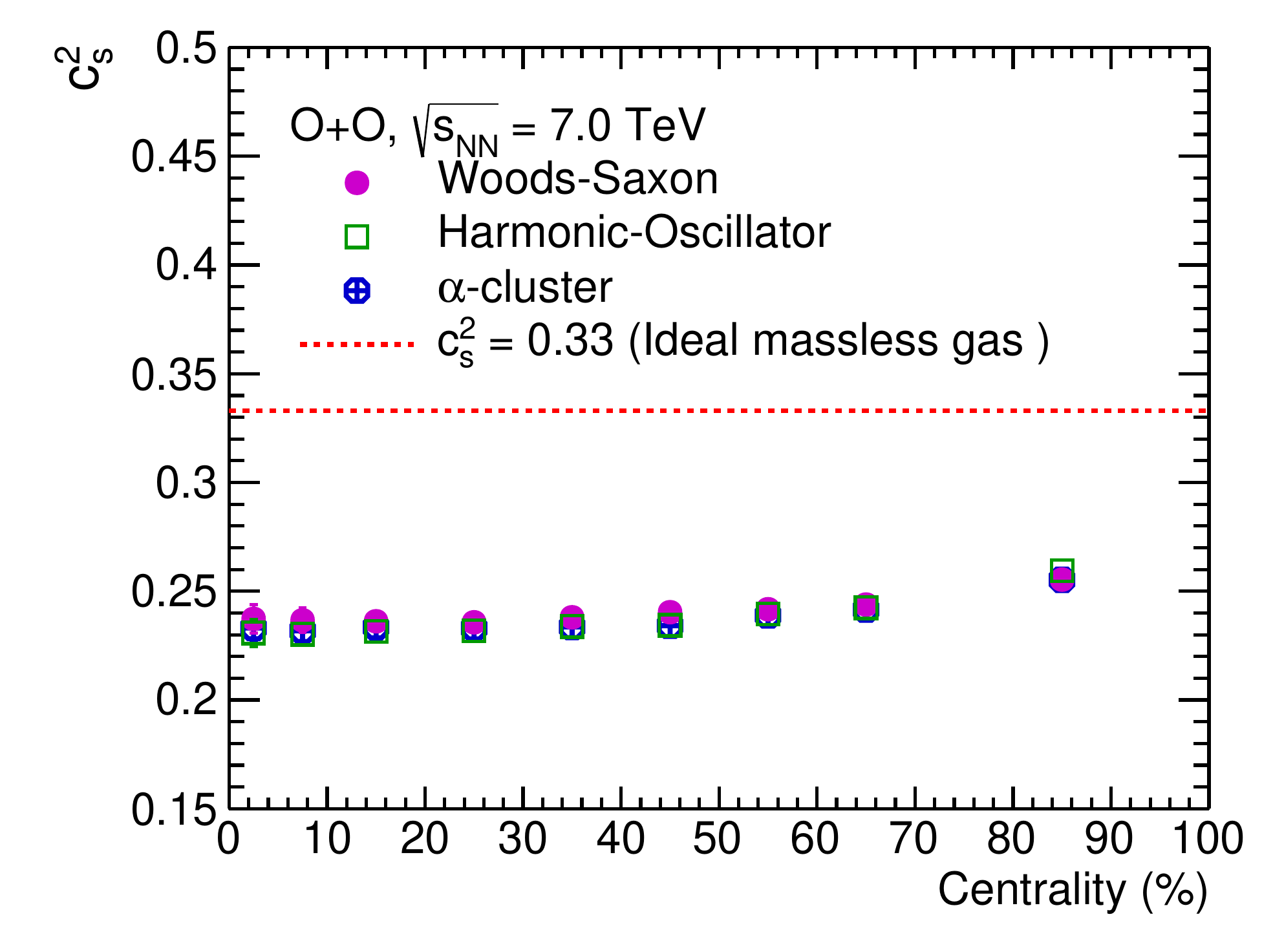}
\caption[]{(Color online) Squared speed of sound as a function centrality classes for pions, kaons and protons in O+O collisions at $\sqrt{s_{\rm{NN}}}$ = 7 TeV. The solid (open) markers represent the Woods-Saxon (harmonic oscillator) potential, and the markers with a cross represent the $\alpha$-clustered structure.}
\label{fig6}
\end{figure}


\begin{figure*}[ht!]
\includegraphics[scale=0.40]{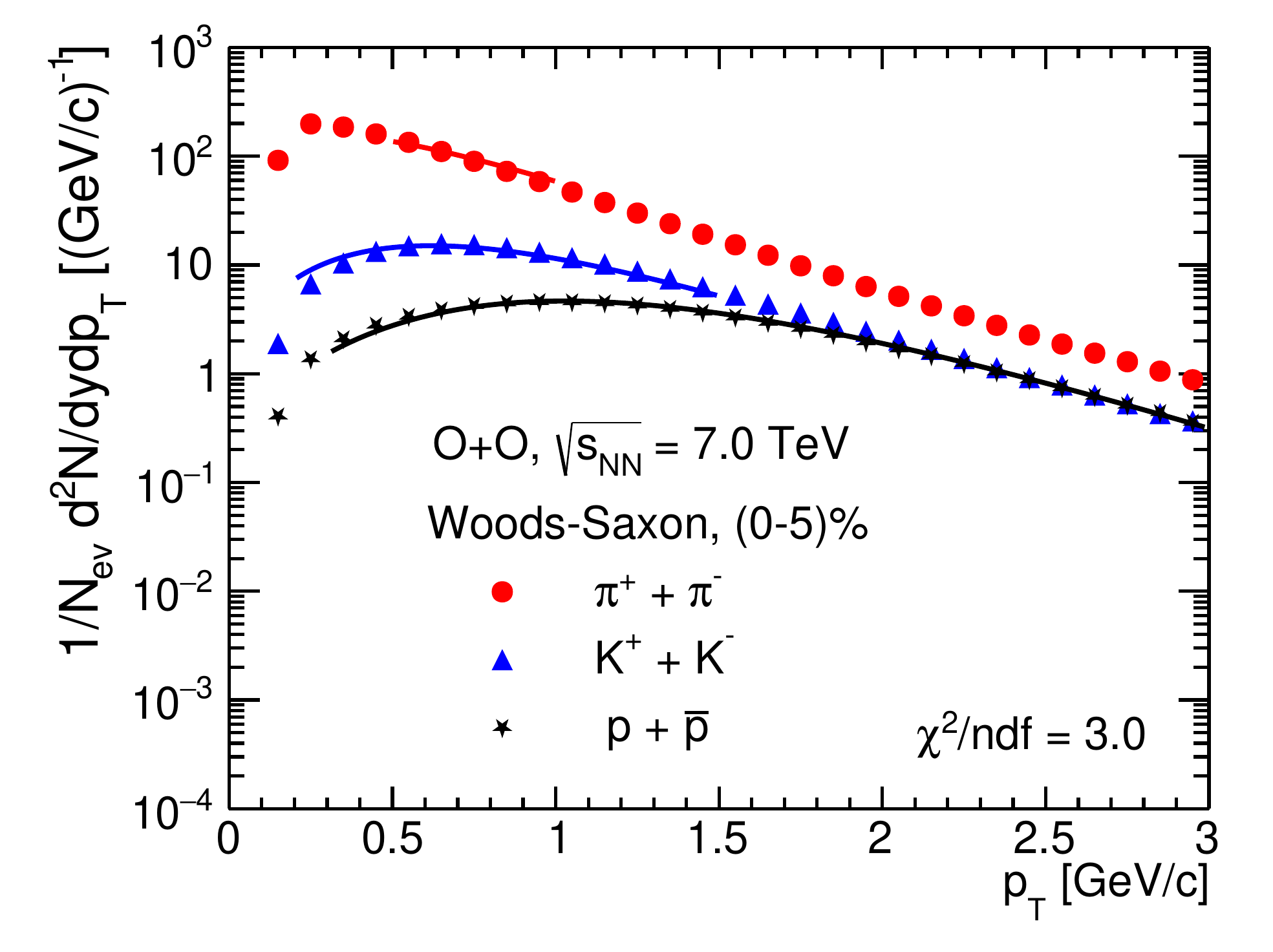}
\includegraphics[scale=0.40]{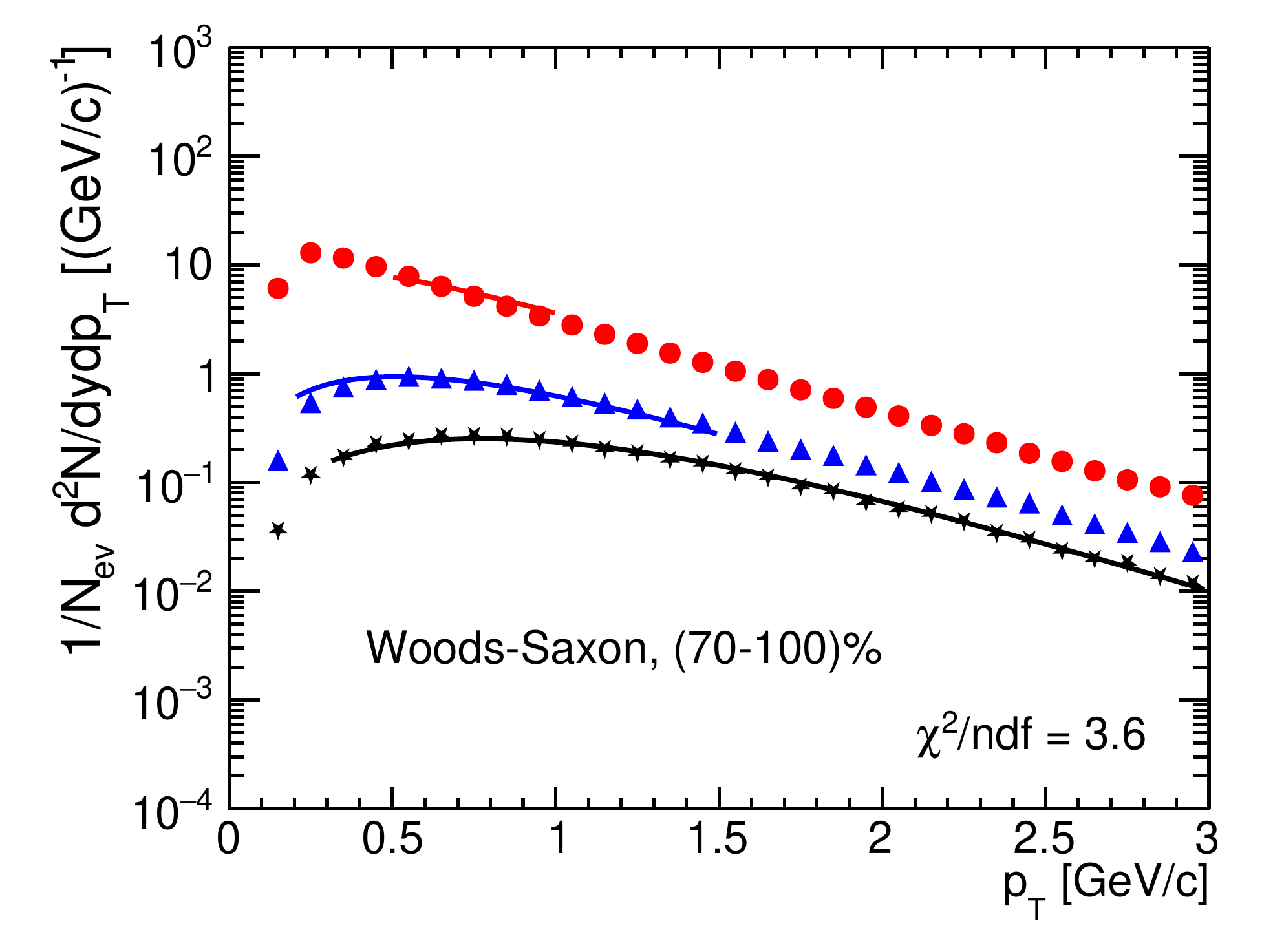}
\includegraphics[scale=0.40]{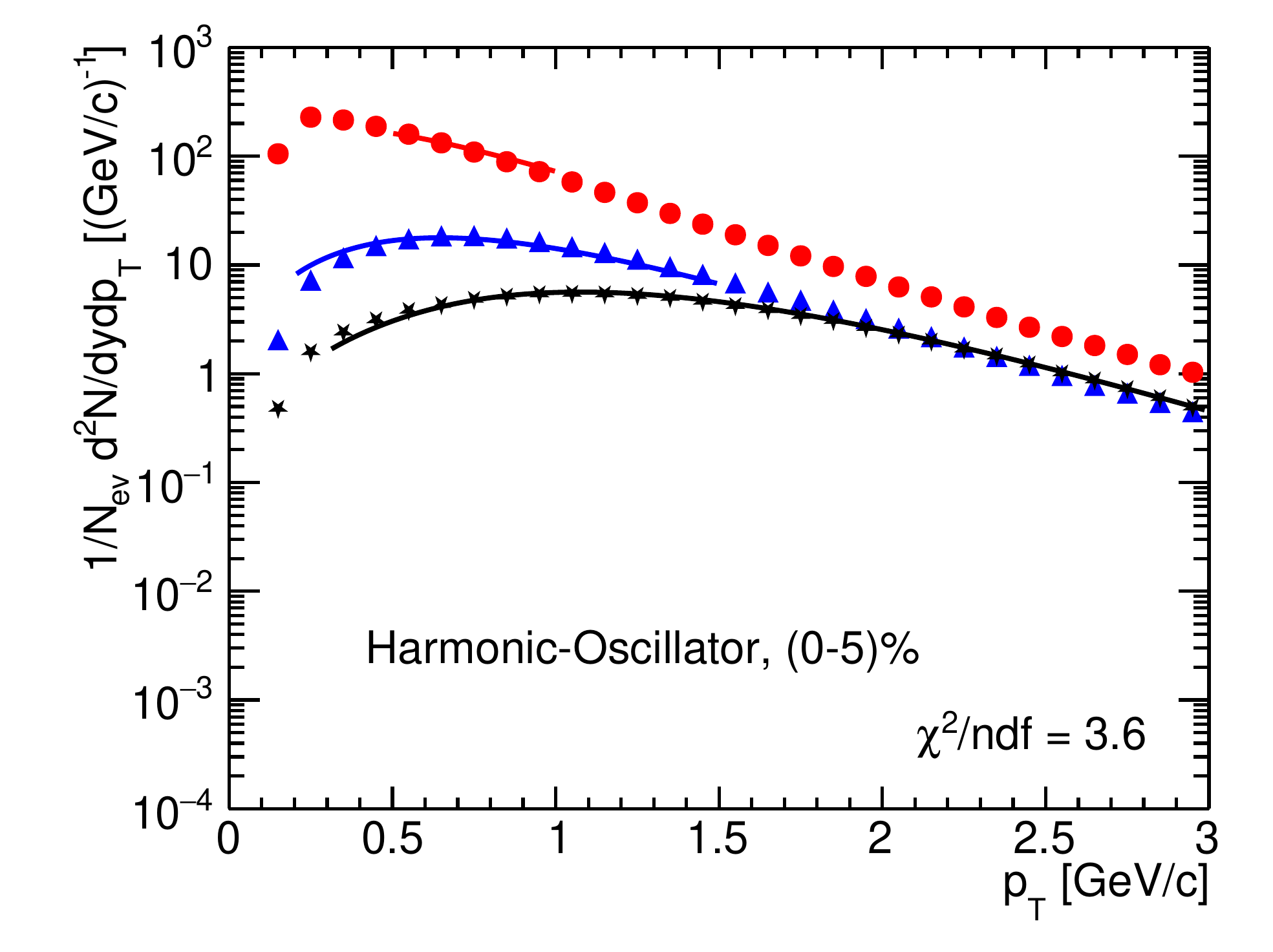}
\includegraphics[scale=0.40]{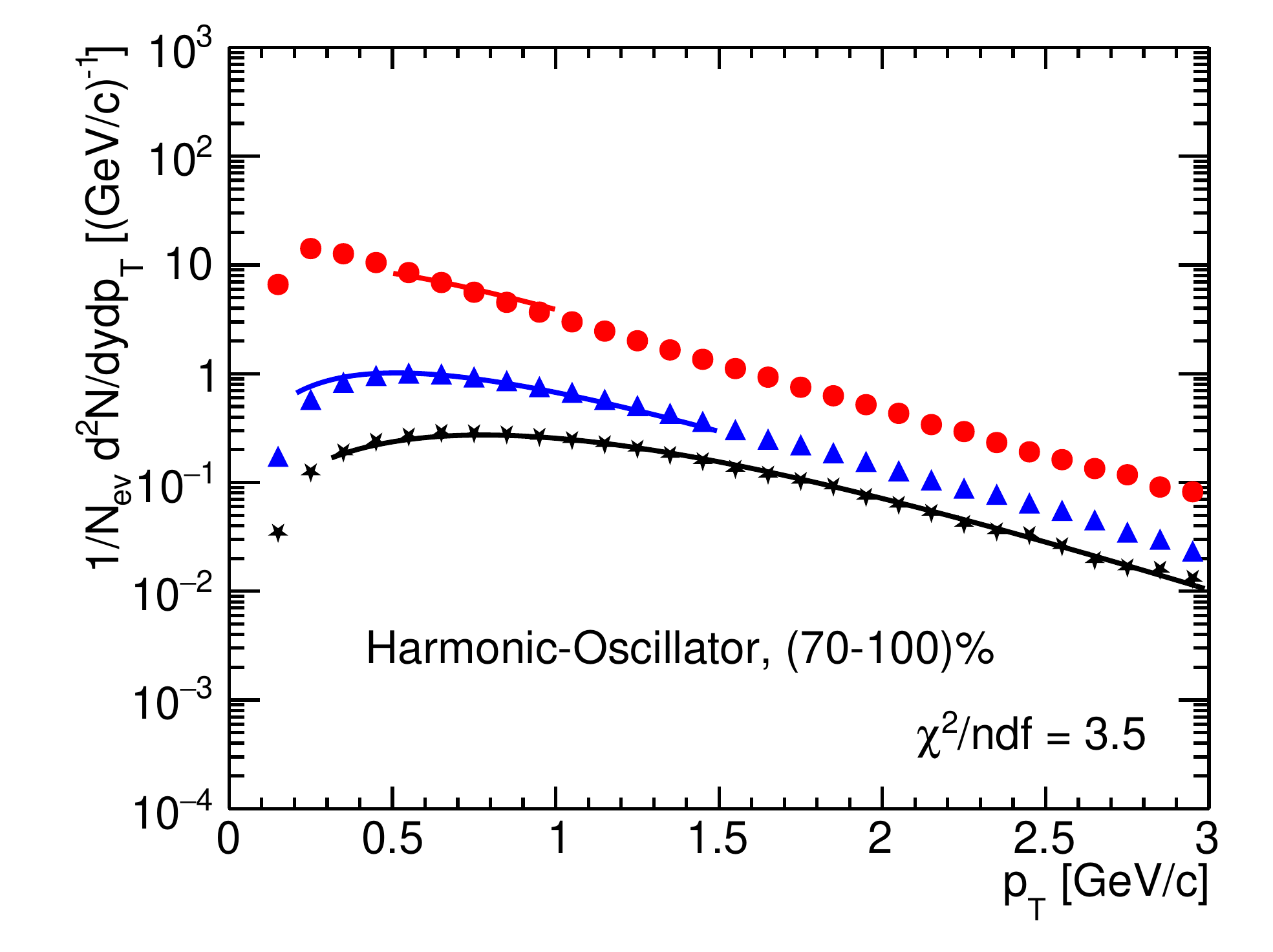}
\includegraphics[scale=0.40]{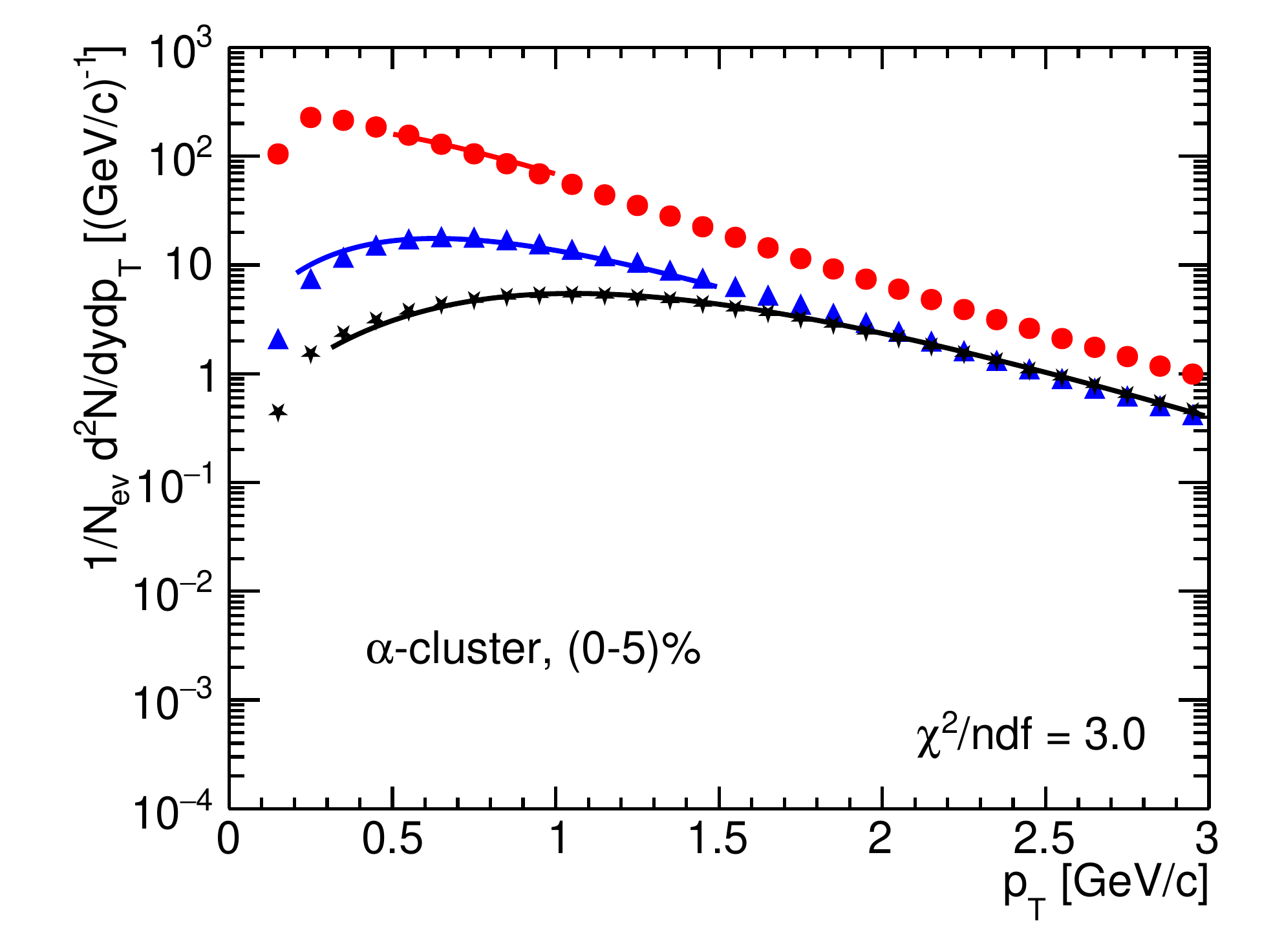}
\includegraphics[scale=0.40]{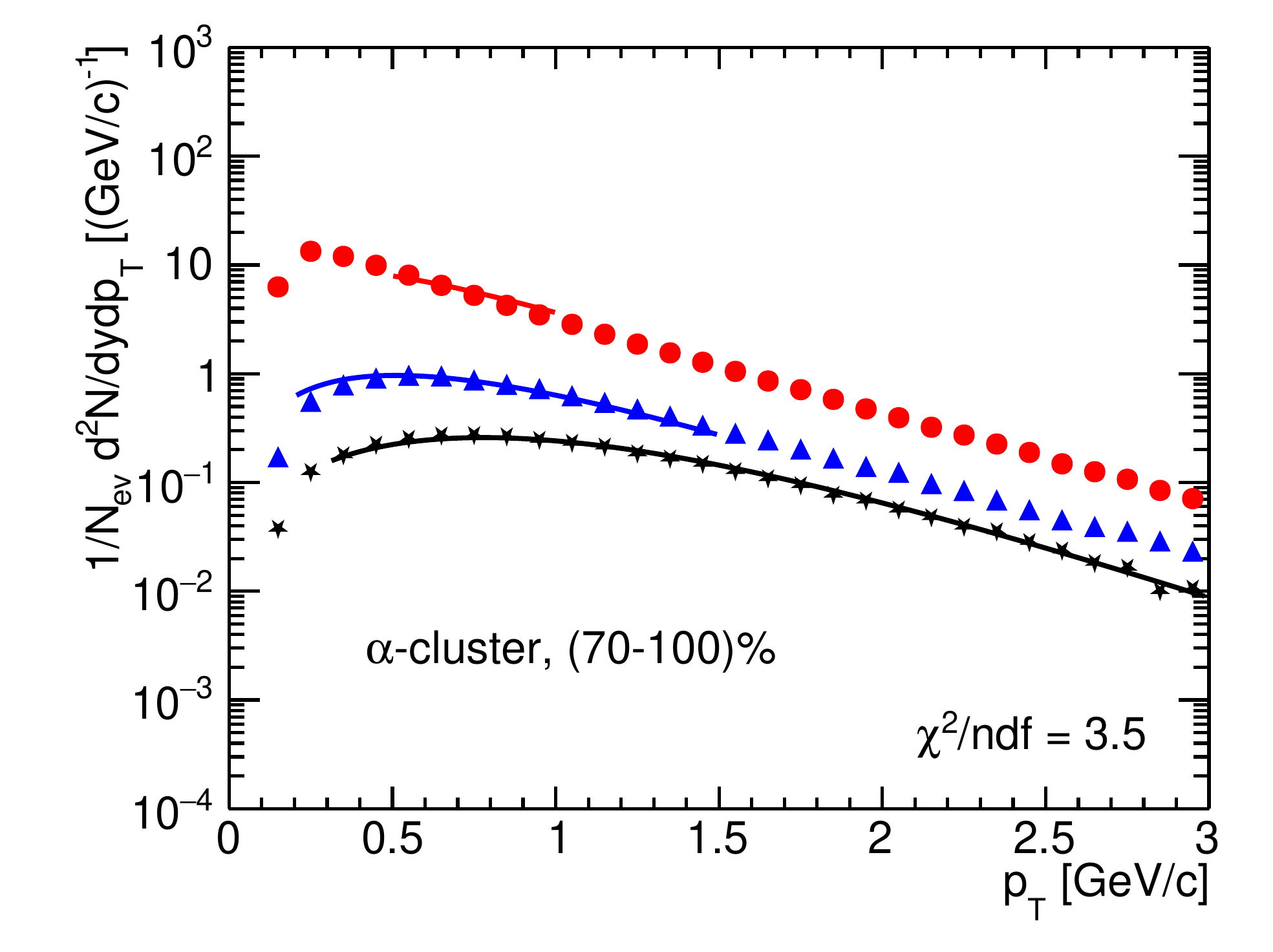}
\caption[]{(Color online) Simultaneous fitting of identified particles' $p_{\rm T}$-spectra with Boltzmann-Gibbs blastwave distribution in  O+O collisions at $\sqrt{s_{\rm{NN}}}$ = 7 TeV. The left and right plots show the fitting  for (0-5)\% and (70-100)\% in the oxygen nucleus for all cases.}
\label{fig8}
\end{figure*}

In Landau hydrodynamical model~\cite{Landau:1953gs}, the rapidity distributions are expected to follow a Gaussian distribution and in this framework, the speed of sound ($c_s$) is related to the width of the rapidity distribution via the following expression.
\begin{equation}
\sigma_{y}^2 = \frac{8}{3}\frac{c_s^2}{1-c_s^2}\ln\bigg(\frac{\sqrt{s_{\rm{NN}}}}{2m_p}\bigg).
\label{eq-cs2}
\end{equation}
Here, $\sigma_{y}$ is the width of the rapidity distribution and $m_p$ is the mass of a proton. The massless ideal gas limit for the squared speed of sound ($c_{s}^2$) is 0.33.  However, the presence of the dip structure in pseudorapidity distributions at around $|\eta| = 0$ makes it difficult to fit via a single Gaussian distribution. Usually, in experiments~\cite{ALICE:2013jfw,RSahoo:2014},  the following double Gaussian distribution is used to fit the pseudorapidity distributions to extrapolate the distributions to the unmeasured regions:
\begin{equation}
dN/{d\eta} = A_{1}e^{\frac{-\eta^2}{2\sigma_1^{2}}} - A_{2}e^{\frac{-\eta^2}{2\sigma_2^{2}}}.
\label{eq-doublegaus}
\end{equation}
Here, $A_{1}$ and $A_{2}$ are the normalization parameters while $\sigma_1$ and $\sigma_2$ are the widths of each Gaussian distribution. Thus, using Eq.~\ref{eq-doublegaus} for the fitting of pseudorapidity distributions, we have obtained $\sigma_1$ and $\sigma_2$. The values of $\sigma_1$ and $\sigma_2$ are found to be similar within uncertainties, which can be seen in Table~\ref{tab:sigma}. Figure~\ref{fig6} shows the squared speed of sound as a function centrality classes for pions, kaons and protons in O+O collisions at $\sqrt{s_{\rm{NN}}}$ = 7 TeV for Woods-Saxon, harmonic oscillator density profiles and $\alpha$-clustered structure obtained using Eq.~\ref{eq-cs2}. Here, $\sigma_1$ has been used to obtain the absolute value and the maximum deviation of $\sigma_1$ and $\sigma_2$ is used as uncertainties for $c_{s}^2$. Considering the uncertainties, $c_{s}^2$ is found to be similar as a function of centrality. The negligible dependence of $c_{s}^2$ on centrality classes could give an indication that the system produced in O+O collisions is significantly less dense compared Pb+Pb collisions and they are more similar to pp collisions. The ideal gas limit is also shown as a red dotted line in Fig.~\ref{fig6} and the observed values of $c_{s}^2$ is found to be around 27\% lower than the massless ideal gas limit.

\begin{table} [!hpt]
                \centering
                \caption{ Double-Gaussian width parameters from fitting the pseudorapidity distributions in the range $|\eta|<3$ using Eq.~\ref{eq-doublegaus}. \label{tab:sigma}}
                \scalebox{0.60}{
                \begin{tabular}{|c |c |c | c| c |c |c |}
                \hline
                \textbf{Centrality(\%)} & \multicolumn{2}{c|}{{\textbf{Woods-Saxon}}} & \multicolumn{2}{c|}{{\textbf{harmonic oscillator}}} & \multicolumn{2}{c|}{{\textbf{$\alpha$-cluster}}}\\ 
                \hline
                & $\sigma_{1}$ & $\sigma_{2}$ & $\sigma_{1}$ & $\sigma_{2}$ & $\sigma_{1}$ & $\sigma_{2}$\\
                \hline
             	\hline
                0--5             &2.35 $\pm$ 0.04       &2.22 $\pm$ 0.04    &2.31 $\pm$ 0.04  &2.20 $\pm$ 0.04            &2.33 $\pm$ 0.04  &2.21 $\pm$ 0.04 \\
                \hline
                5--10           &2.34 $\pm$ 0.04       &2.24 $\pm$ 0.04       &2.31 $\pm$ 0.04  &2.24 $\pm$ 0.03         &2.32 $\pm$ 0.04  &2.22 $\pm$ 0.04 \\
                \hline
                10--20           &2.34 $\pm$ 0.03       &2.25 $\pm$ 0.03    &2.32 $\pm$ 0.04  &2.21 $\pm$ 0.03          &2.33 $\pm$ 0.04  &2.22 $\pm$ 0.04 \\
                \hline
                20--30           &2.34 $\pm$ 0.02       &2.27 $\pm$ 0.03       &2.32 $\pm$ 0.03  &2.24 $\pm$ 0.03       &2.32 $\pm$ 0.03  &2.25 $\pm$ 0.03 \\
                \hline
                30--40           &2.35 $\pm$ 0.02       &2.30 $\pm$ 0.02       &2.33 $\pm$ 0.02  &2.27 $\pm$ 0.02        &2.33 $\pm$ 0.02  &2.27 $\pm$ 0.02 \\
                \hline
                40--50           &2.36 $\pm$ 0.01       &2.33 $\pm$ 0.01       &2.33 $\pm$ 0.02  &2.29 $\pm$ 0.02        &2.33 $\pm$ 0.02  &2.29 $\pm$ 0.02  \\
                \hline
                50--60            &2.37 $\pm$ 0.01       &2.35 $\pm$ 0.01       &2.36 $\pm$ 0.01  &2.33 $\pm$ 0.01       &2.36 $\pm$ 0.01  &2.32 $\pm$ 0.01 \\
                \hline
                60--70            &2.38 $\pm$ 0.01       &2.36 $\pm$ 0.01       &2.38 $\pm$ 0.01  &2.35 $\pm$ 0.01       &2.37 $\pm$ 0.01  &2.35 $\pm$ 0.01 \\
                \hline
                70--100          &2.48 $\pm$ 0.01       &2.44 $\pm$ 0.01       &2.47 $\pm$ 0.01  &2.46 $\pm$ 0.01       &2.45 $\pm$ 0.01  &2.44 $\pm$ 0.01 \\
                \hline
                \end{tabular} 
                }              
\end{table}

\subsection{$p_{\rm T}$-spectra and kinetic freeze-out parameters}

\begin{figure}[ht]
\includegraphics[scale=0.40]{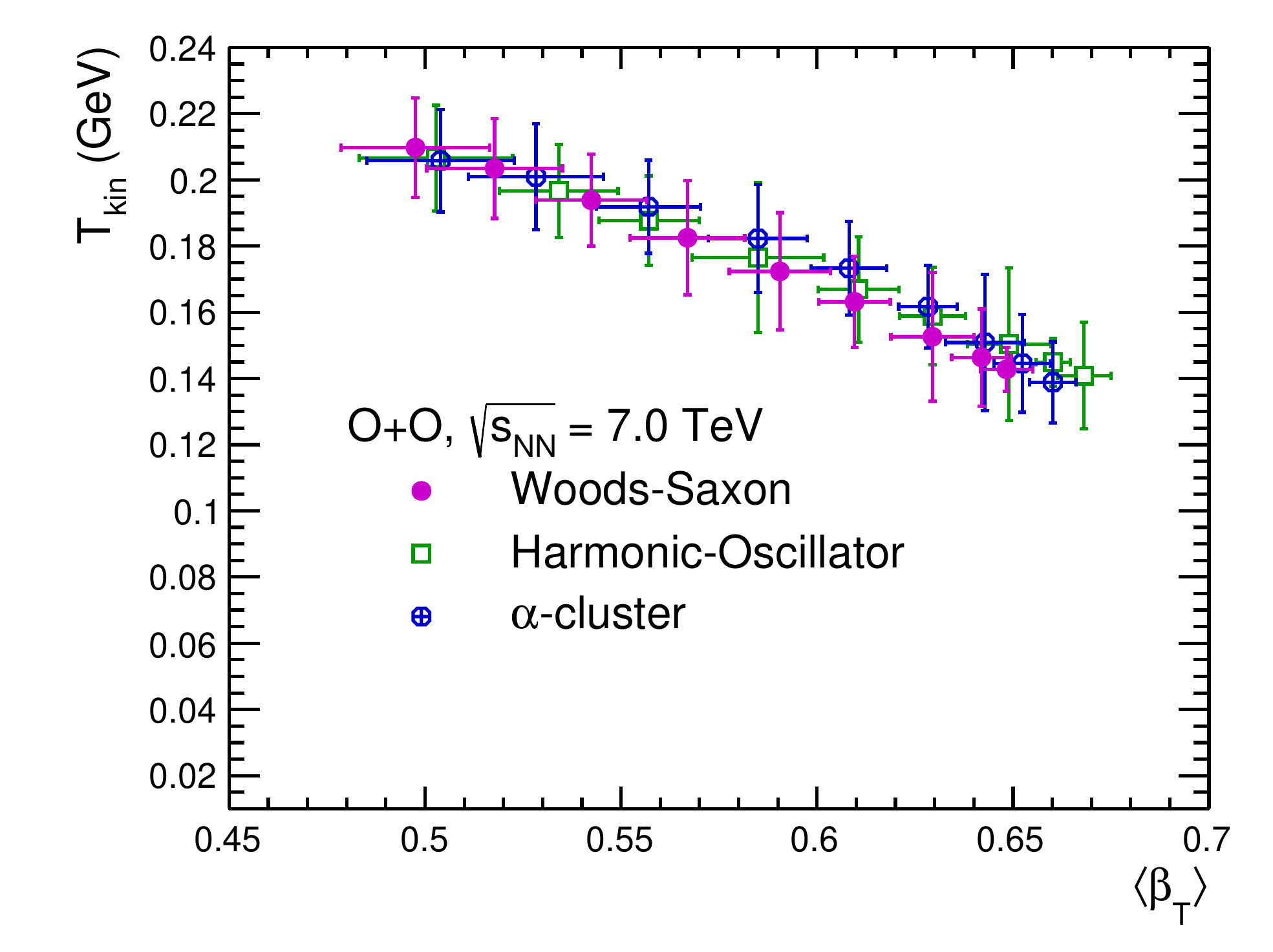}
\caption[]{(Color online) Kinetic freeze-out temperature versus transverse radial flow from simultaneous fit of identified particles' $p_{\rm T}$-spectra with Boltzmann-Gibbs blastwave distribution in  O+O collisions at $\sqrt{s_{\rm{NN}}}$ = 7 TeV. The solid (open) markers represent the Woods-Saxon (harmonic oscillator) density profile, and the markers with a cross represent the $\alpha$-clustered structure.}
\label{fig9}
\end{figure}

Figure~\ref{fig8} shows the simultaneous fitting of identified particles' $p_{\rm T}$-spectra with Boltzmann-Gibbs blastwave distribution in  O+O collisions at $\sqrt{s_{\rm{NN}}}$ = 7 TeV for (0-5)\% (left) and (70-100)\% (right) centrality classes for Woods-Saxon, harmonic oscillator density profiles and $\alpha$-clustered structure in oxygen nucleus. The fitting ranges for each particle are similar to the range reported by ALICE~\cite{ALICE:2019hno}. The fitting is performed via the $\chi^2$-minimisation method and the values of $\chi^2$ per degree of freedom are shown in each of the cases in Fig~\ref{fig8}. 
The expression for invariant yield in the Boltzmann-Gibbs blastwave framework~\cite{Schnedermann:1993ws} is given as the following:
 \begin{eqnarray}
\label{bgbw1}
E\frac{d^3N}{dp^3}=C \int d^3\sigma_\mu p^\mu \exp(-\frac{p^\mu u_\mu}{T_{\rm{kin}}}).
\end{eqnarray}
Here $C$ is the normalisation constant. The particle four-momentum is given by, 
 \begin{eqnarray}
p^\mu~=~(m_T{\cosh}y,~p_T\cos\phi,~ p_T\sin\phi,~ m_T{\sinh}y),
\end{eqnarray}
and the four-velocity is given by,
 \begin{eqnarray}
u^\mu=\cosh\rho~(\cosh\eta,~\tanh\rho~\cos\phi_r,~\tanh\rho~\sin~\nonumber\\
\phi_r,~\sinh~\eta).
\end{eqnarray}
Finally, the freeze-out surface is parametrised as, 
 \begin{eqnarray}
d^3\sigma_\mu~=~(\cosh\eta,~0,~0, -\sinh\eta)~\tau~r~dr~d\eta~d\phi_r.
\end{eqnarray}
here, $\eta$ is the space-time rapidity. Now, Eq.~\ref{bgbw1} is expressed as,
 \begin{eqnarray}
\label{boltz_blast}
\left.\frac{d^2N}{dp_Tdy}\right|_{y=0} = Cp_{T}m_{T} \int_0^{R_{0}} r\;dr\;K_1\Big(\frac{m_T\;\cosh\rho}{T_{\rm{kin}}}\Big)I_0\nonumber\\
\Big(\frac{p_T\;\sinh\rho}{T_{\rm{kin}}}\Big).
\end{eqnarray}
 $K_{1}\displaystyle\Big(\frac{m_T\;{\cosh}\rho}{T_{\rm{kin}}}\Big)$ and $I_0\displaystyle\Big(\frac{p_T\;{\sinh}\rho}{T_{\rm{kin}}}\Big)$ are modified Bessel's functions, which are given by,
 \begin{eqnarray}
\centering
K_1\Big(\frac{m_T\;{\cosh}\rho}{T_{\rm{kin}}}\Big)=\int_0^{\infty} {\cosh}y\;{\exp} \Big(-\frac{m_T\;{\cosh}y\;{\cosh}\rho}{T_{\rm{kin}}}\Big)dy\nonumber,
\end{eqnarray}
 \begin{eqnarray}
\centering
I_0\Big(\frac{p_T\;{\sinh}\rho}{T_{\rm{kin}}}\Big)=\frac{1}{2\pi}\int_0^{2\pi} \exp\Big(\frac{p_T\;{\sinh}\rho\;{\cos}\phi}{T_{\rm{kin}}}\Big)d\phi \nonumber,
\end{eqnarray}
where, $\rho$ is given by $\rho={\tanh}^{-1}\beta_{\rm T}$ and $\beta_{\rm T}(=\displaystyle\beta_s\xi^n$) \cite{Schnedermann:1993ws,Huovinen:2001cy,BraunMunzinger:1994xr, Tang:2011xq} is the radial flow velocity. Here, $\xi$ is given as $\displaystyle(r/R_0)$, $\beta_s$ is the maximum surface velocity and $r$ is the radial distance. $R_0$ is the maximum radius of the fireball at freeze-out. In this model, the particles closer to the center of the fireball are assumed to move slower than the ones at the edges. The average of the transverse velocity is evaluated as \cite{Adcox:2003nr}, 
 \begin{eqnarray}
\langle\beta_{\rm T}\rangle =\frac{\int \beta_s\xi^n\xi\;d\xi}{\int \xi\;d\xi}=\Big(\frac{2}{2+n}\Big)\beta_s.
\end{eqnarray}
For our calculation, we use $n$ as a free parameter. Figure~\ref{fig9} shows the kinetic freeze-out temperature versus transverse radial flow velocity from the simultaneous fit of identified particles' $p_{\rm T}$-spectra with Boltzmann-Gibbs blastwave distribution (Eq.~\ref{boltz_blast}) in O+O collisions at $\sqrt{s_{\rm{NN}}}$ = 7 TeV. The solid (open) markers represent the Woods-Saxon (harmonic oscillator) density profile and the markers with a cross represent $\alpha$-clustered structure nucleus. Within uncertainties, the correlation between kinetic freeze-out temperature ($T_{\rm{kin}}$) and average transverse flow ($\langle\beta_{T}\rangle$) is similar for all cases. For the most central collisions (0-5\% class), $T_{\rm{kin}}$ is the lowest and the transverse flow is the highest. This behavior is expected as the most central collisions have the largest system size due to which the hadronic phase lasts longer, which makes the $T_{\rm{kin}}$ lowest. Also, due to the 
largest system size, the radial flow is expected to be the highest. A similar behavior is seen for Pb+Pb collisions at LHC energies~\cite{ALICE:2019hno}.

\begin{table} [!hpt]
                \centering
                \caption{Kinetic freeze-out temperature and transverse radial flow parameter obtained from Boltzmann-Gibbs blastwave fit using Eq.~\ref{boltz_blast}.}
                \scalebox{0.57}{
                \begin{tabular}{|c |c |c | c| c | c| c|}
                \hline
                \textbf{Centrality(\%)} & \multicolumn{2}{c|}{{\textbf{Woods-Saxon}}} & \multicolumn{2}{c|}{{\textbf{harmonic oscillator}}} & \multicolumn{2}{c|}{{\textbf{$\alpha$-cluster}}}\\ 
                \hline
                & $T_{\rm{kin}}$ (GeV) & $\langle\beta_{T}\rangle$ & $T_{\rm{kin}}$ (GeV) & $\langle\beta_{T}\rangle$ & $T_{\rm{kin}}$ (GeV) & $\langle\beta_{T}\rangle$\\
                \hline
             	\hline
                0--5             &0.143 $\pm$ 0.013       &0.65 $\pm$ 0.01    &0.141 $\pm$ 0.016  &0.67 $\pm$ 0.01                &0.139 $\pm$ 0.012  &0.66 $\pm$ 0.01 \\
                \hline
                5--10           &0.146 $\pm$ 0.015       &0.64 $\pm$ 0.01       &0.145 $\pm$ 0.007  &0.66 $\pm$ 0.00             &0.145 $\pm$ 0.015  &0.65 $\pm$ 0.01\\
                \hline
                10--20           &0.153 $\pm$ 0.019       &0.63 $\pm$ 0.01    &0.150 $\pm$ 0.231  &0.65 $\pm$ 0.01              &0.151 $\pm$ 0.021  &0.64 $\pm$ 0.01\\
                \hline 
                20--30           &0.163 $\pm$ 0.014       &0.61 $\pm$ 0.01       &0.159 $\pm$ 0.015  &0.63 $\pm$ 0.01           &0.162 $\pm$ 0.012  &0.63 $\pm$ 0.01\\
                \hline
                30--40           &0.172 $\pm$ 0.018       &0.59 $\pm$ 0.01       &0.167 $\pm$ 0.016  &0.61 $\pm$ 0.01           &0.173  $\pm$ 0.014  &0.61 $\pm$ 0.01\\
                \hline
                40--50           &0.182 $\pm$ 0.017       &0.57 $\pm$ 0.01      &0.177 $\pm$ 0.022  &0.58 $\pm$ 0.02          &0.182 $\pm$ 0.016  &0.58 $\pm$ 0.01\\
                \hline
                50--60            &0.194 $\pm$ 0.013       &0.54 $\pm$ 0.01       &0.188 $\pm$ 0.013  &0.56 $\pm$ 0.01          &0.192 $\pm$ 0.014  &0.56 $\pm$ 0.01\\
                \hline
                60--70            &0.203 $\pm$ 0.015       &0.52 $\pm$ 0.02       &0.197 $\pm$ 0.014  &0.53 $\pm$ 0.01          &0.201 $\pm$ 0.016  &0.53 $\pm$ 0.02\\
                \hline
                70--100          &0.210 $\pm$ 0.015       &0.50 $\pm$ 0.02       &0.206 $\pm$ 0.008  &0.50 $\pm$ 0.01         &0.206 $\pm$ 0.016  &0.50 $\pm$ 0.02\\
                \hline
                \end{tabular} 
                }              
\end{table}

\section{Particle ratios}
\begin{figure*}[ht]
\includegraphics[scale=0.40]{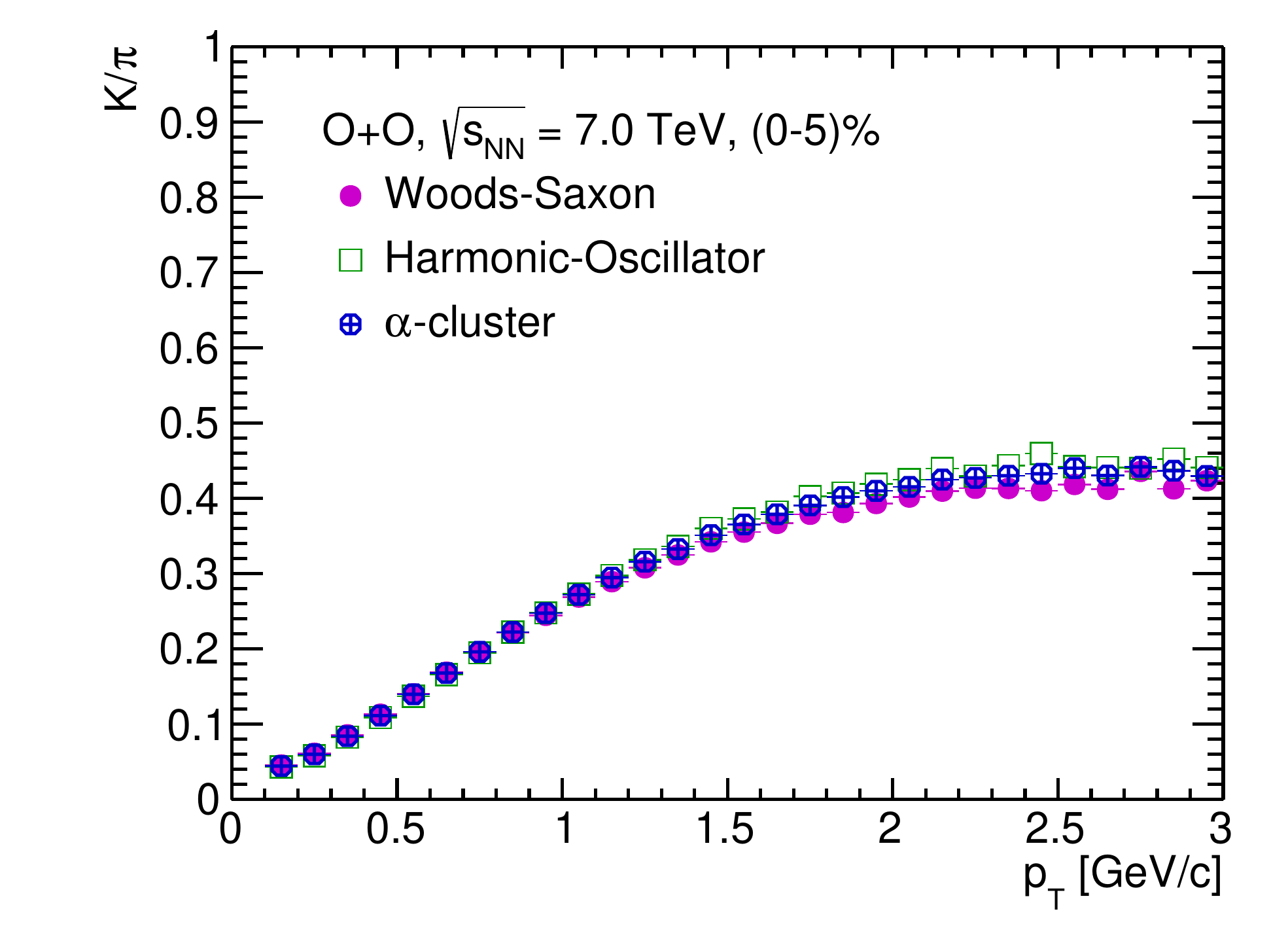}
\includegraphics[scale=0.40]{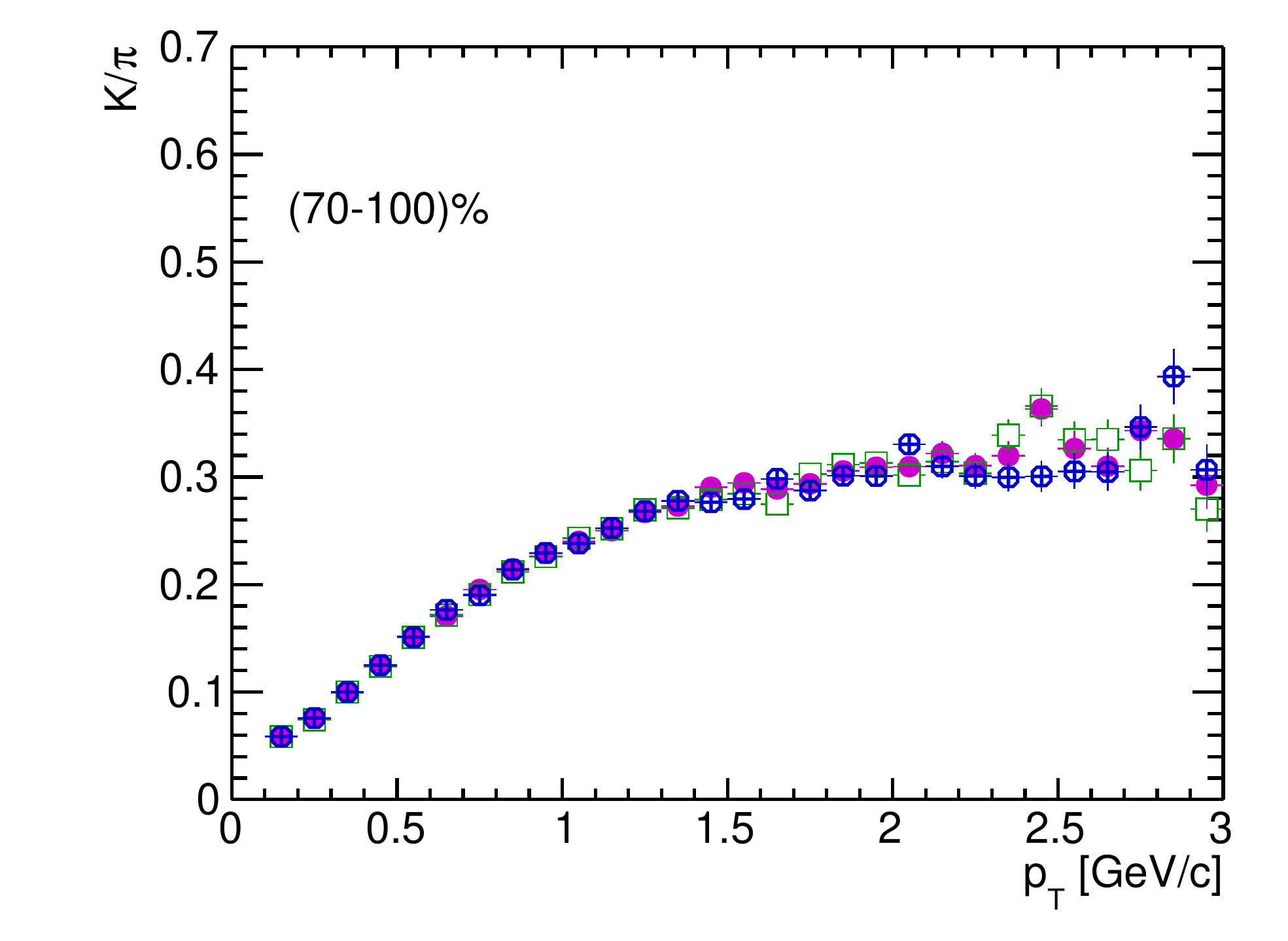}
\includegraphics[scale=0.40]{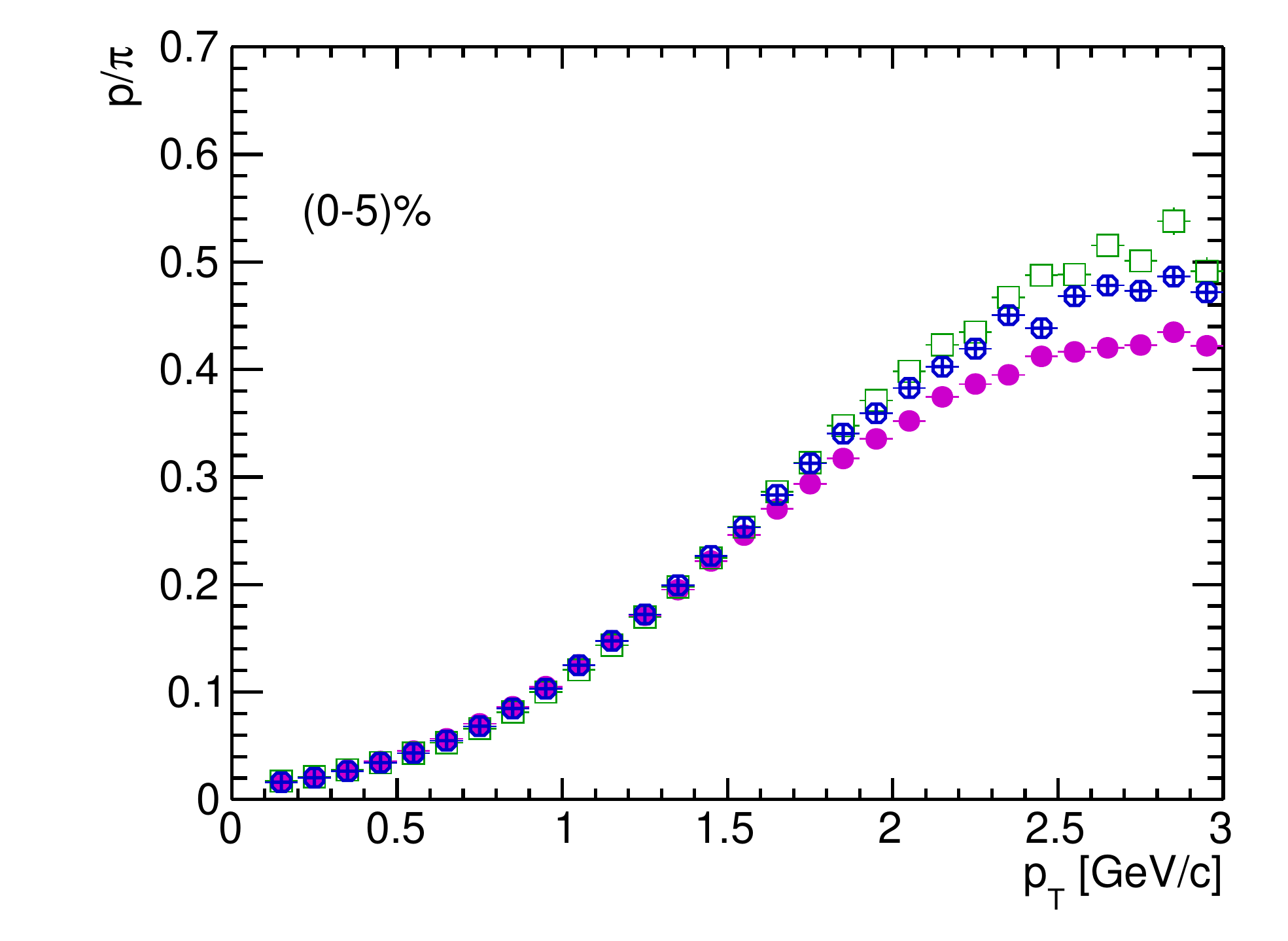}
\includegraphics[scale=0.40]{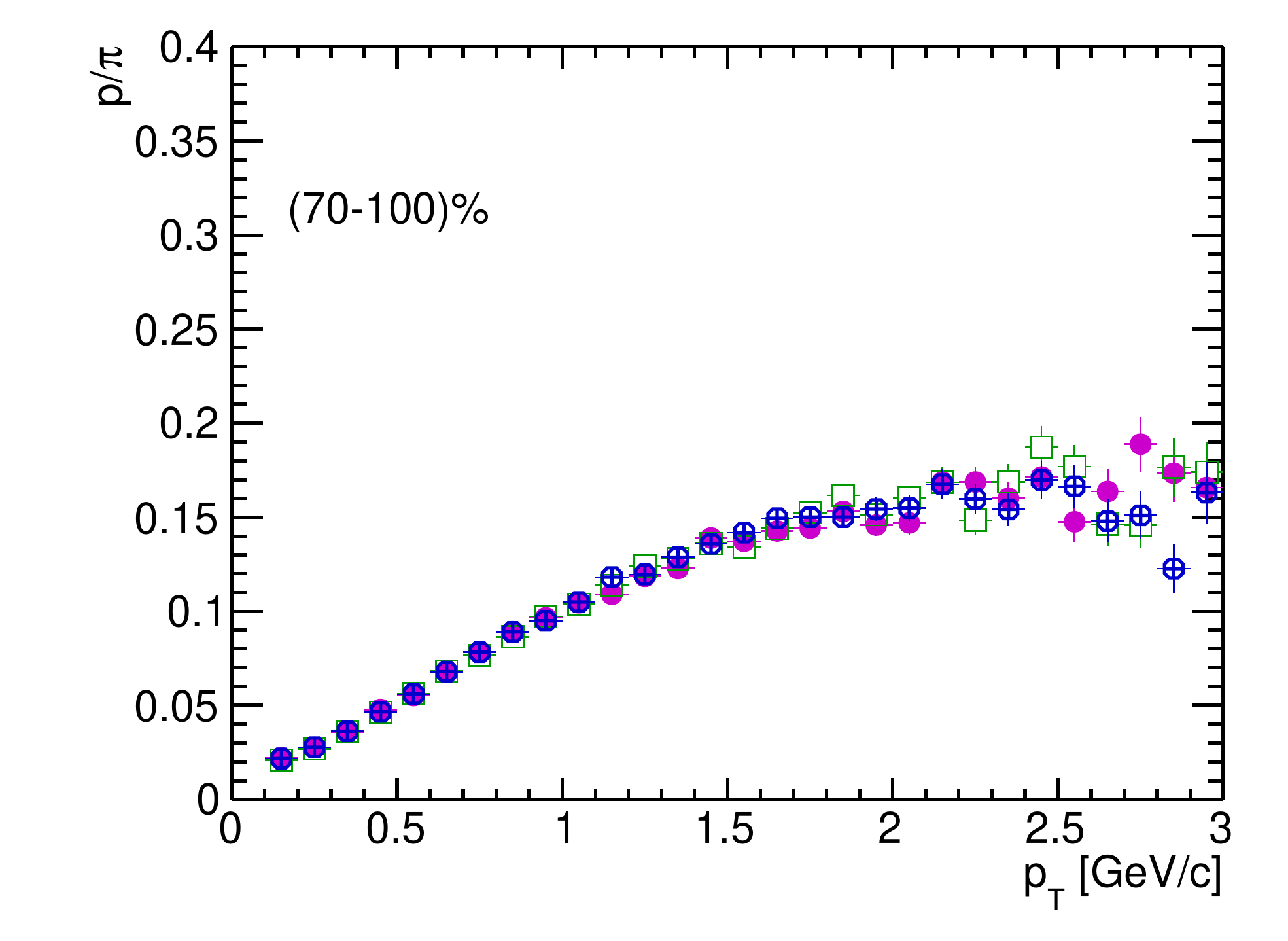}
\caption[]{(Color online)$p_{\rm T}$-differential kaon-to-pion (top) and proton-to-pion (bottom) ratio in  O+O collisions at $\sqrt{s_{\rm{NN}}}$ = 7 TeV for (0-5)\% (left) and (70-100)\% (right) centrality classes. The solid (open) markers represent the Woods-Saxon (harmonic oscillator) density profile and the markers with a cross represent $\alpha$-clustered  structure of oxygen nucleus.}
\label{fig10}
\end{figure*}

Figure~\ref{fig10} shows $p_{\rm T}$-differential kaon-to-pion and proton-to-pion ratios in  O+O collisions at $\sqrt{s_{\rm{NN}}}$ = 7 TeV for (0-5)\% and (70-100)\% centrality classes. Both the ratios to pions increase as a function of $p_{\rm T}$. As the kaon-to-pion ratio is the measure of strangeness, we see enhancement of strangeness production as a function of $p_{\rm T}$. This enhancement seems to be the similar for both (0-5)\% and (70-100)\% centrality classes at low-$p_{\rm T}$. However, the enhancement is higher for (0-5)\% centrality class at intermediate-$p_{\rm T}$. The proton-to-pion ratio, a ratio between the lightest baryon to lightest meson, acts as a proxy for the baryon to meson ratio. In general, the particle ratios do not show any dependence on the density profiles for (70-100)\%. However, for (0-5)\% centrality class, a rise in both kaon-to-pion and proton-to-pion ratios is seen at intermediate-$p_{\rm T}$ for harmonic oscillator density profile with respect to Woods-Saxon density profile, which becomes prominent at higher $p_{\rm T}$. The results for the $\alpha$-clustred structure are similar to that observed for the harmonic oscillator density profile.

\section{Elliptic flow}
The initial spatial anisotropies of the overlap region in non-central heavy-ion collisions get converted to the momentum space azimuthal anisotropy of the final state particles due to the differential pressure gradients of the produced medium. The azimuthal anisotropy can be expressed as a Fourier series in the azimuthal angle, $\phi$:
\begin{eqnarray}
E\frac{d^3N}{dp^3}=\frac{d^2N}{2\pi p_{\rm T}dp_{\rm T}dy}\bigg(1+2\sum_{n=1}^\infty v_n \cos[n(\phi -\psi_n)]\bigg)\,,\nonumber\\
\label{eq2}
\end{eqnarray}
where, $v_n$ is the anisotropic flow of different order $n$, with n = 2 is elliptic flow. $\psi_n$ is the n$^{\text{th}}$ harmonic event plane angle~\cite{v2eventplane}. To reduce the non-flow effects, a two-particle correlation method~\cite{Aad:2015lwa,Mallick:2021rsd} can be adopted to estimate the elliptic flow. The correlation function between two particles is obtained in relative pseudorapidity ($\Delta\eta = \eta_a - \eta_b$) and relative azimuthal angle ($\Delta\phi = \phi_a - \phi_b$). Here, $a$ and $b$ denote two separate particles in a pair. In this study, we have taken charged particles in $|\eta|<$ 2.5 and $p_{\rm T}>$ 0.5 GeV/c to be consistent with previous studies~\cite{Aad:2015lwa,Mallick:2021rsd}. 


The 1D correlation function is given as,
\begin{eqnarray}
C(\Delta\phi)= \frac{dN_{\rm pairs}}{d\Delta\phi} = A \times \frac{\int S(\Delta\eta ,\Delta\phi) d\Delta\eta}{\int B(\Delta\eta ,\Delta\phi)d\Delta\eta},
\label{eq5}
\end{eqnarray}
where, the normalization constant ($A$) ensures that the number of pairs are the same between signal, $S(\Delta\eta ,\Delta\phi)$ and background, $B(\Delta\eta ,\Delta\phi)$. The $\Delta\eta$ interval is chosen carefully, as done in previous studies~\cite{Aad:2015lwa,Mallick:2021rsd}, by excluding the jet peak region observed in the $C(\Delta\eta ,\Delta\phi)$ distribution. The interval is taken to be $2.0< |\Delta\eta| < 4.8$. This pseudorapidity cut removes the residual non-flow effects significantly in the estimation of elliptic flow. These non-flow correlations usually arise from jets and short-range resonance decays, and they are not associated with the anisotropy in the early stage of the collisions. In the current study, the non-flow effects are reduced significantly but they might be still non-zero. Thus, the quantitative interpretation of the results may be taken with caution.

\begin{figure}[ht]
\includegraphics[scale=0.4]{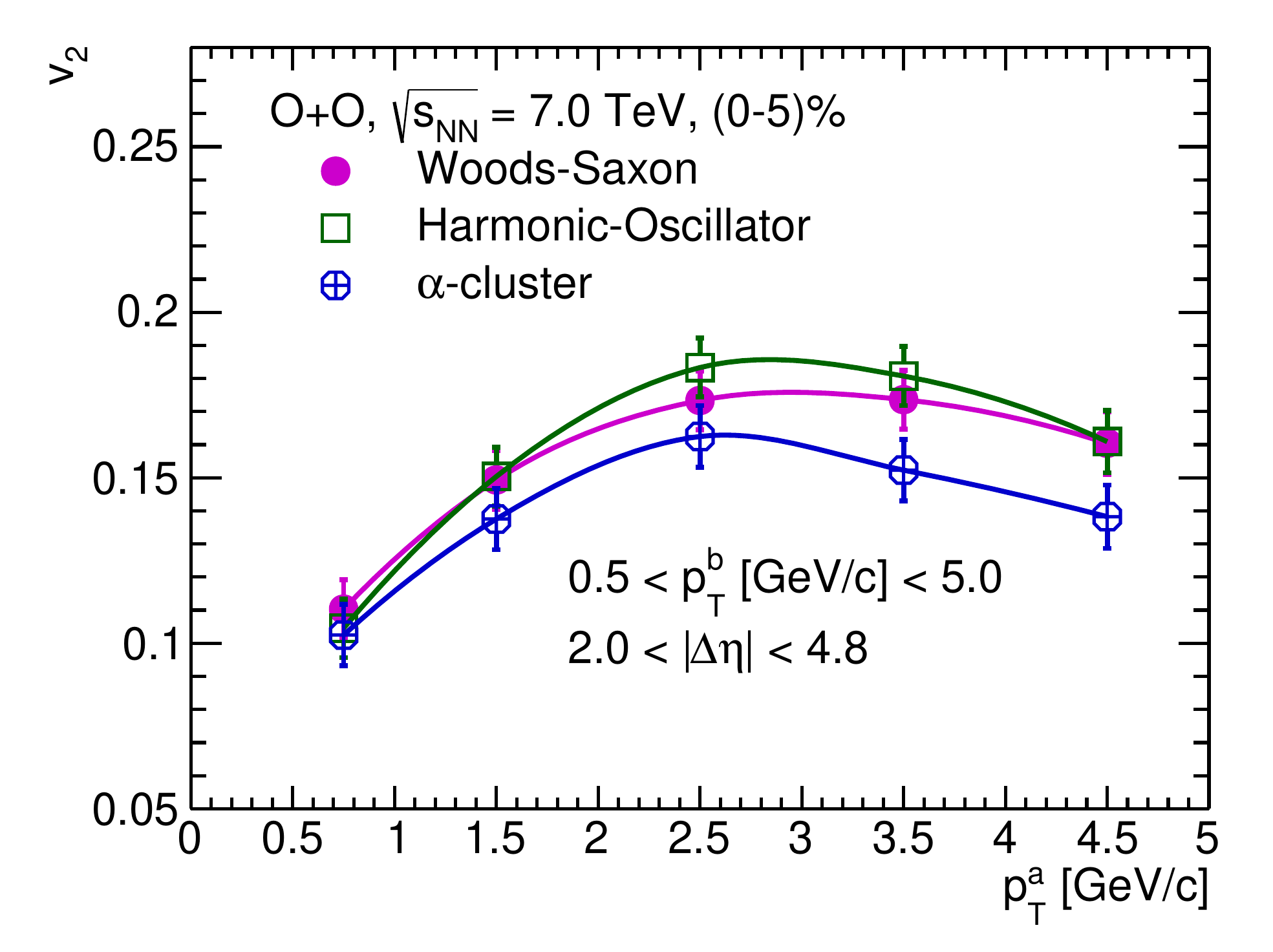}
\includegraphics[scale=0.40]{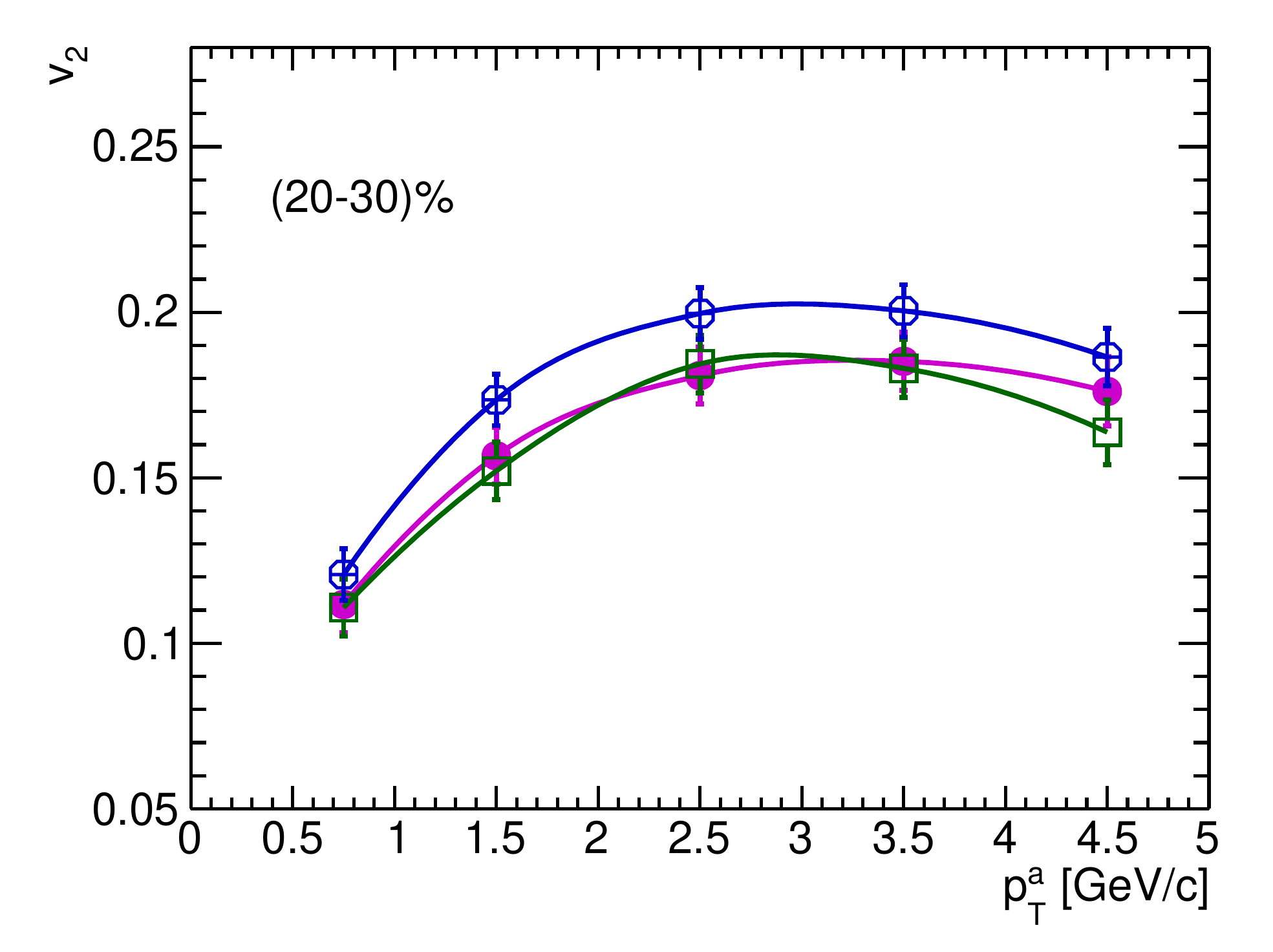}
\caption[]{(Color online)Elliptic flow of charged particles as a function of transverse momentum in O+O collisions for (0-5)\% and (20-30)\%  centrality classes at $\sqrt{s_{\rm{NN}}}$ = 7 TeV. The solid (open) markers represent the Woods-Saxon (harmonic oscillator) density profile of the oxygen nucleus, and the markers with a cross represent the $\alpha$-clustered structure.}
\label{fig11}
\end{figure}

The pair distribution can be expanded into a Fourier series:
\begin{eqnarray}
\frac{dN_{\rm pairs}}{d\Delta\phi} \propto  \bigg[1+2\sum_{n=1}^\infty v_{n,n}(p_{\rm T}^{a},p_{\rm T}^{b}) \cos n\Delta\phi  \bigg],
\label{eq6}
\end{eqnarray}
where, $v_{n,n}$ is the two-particle flow coefficient. Now, Eq.~\ref{eq5} is given as:
\begin{eqnarray}
C(\Delta\phi) \propto \bigg[1+2\sum_{n=1}^\infty v_{n,n}(p_{\rm T}^{a},p_{\rm T}^{b}) \cos n\Delta\phi \bigg].
\label{eq7}
\end{eqnarray}
The definition of harmonics defined in Eq.~\ref{eq2} now enters in Eq.~\ref{eq6},
\begin{eqnarray}
\frac{dN_{\rm pairs}}{d\Delta\phi} \propto  \bigg[ 1+2\sum_{n=1}^\infty v_{n}(p_{\rm T}^{a}) v_{n}(p_{\rm T}^{b}) \cos n\Delta\phi  \bigg].
\label{eq9}
\end{eqnarray}
Through this definition, the event plane angle drops out in convolution. Thus, if the elliptic flow is driven by purely collective expansion then $v_{2,2}$ 
should be factorized into the product of two single-particle elliptic flow coefficients. 
\begin{eqnarray}
v_{2,2}(p_{\rm T}^{a},p_{\rm T}^{b})= v_{2}(p_{\rm T}^{a}) v_{2}(p_{\rm T}^{b}).
\label{eq10}
\end{eqnarray}
Using Eq.~\ref{eq10}, we calculate the single particle elliptic flow coefficient as,
\begin{eqnarray}
v_{2}(p_{\rm T}^{a})= v_{2,2}(p_{\rm T}^{a},p_{\rm T}^{b})/\sqrt{v_{2,2}(p_{\rm T}^{b},p_{\rm T}^{b})}
\label{eq11}
\end{eqnarray}

We now proceed for the estimation of the elliptic flow of charged particles as a function of transverse momentum for (0-5)\% and (20-30)\% centrality classes in O+O collisions at $\sqrt{s_{\rm{NN}}}$ = 7 TeV using AMPT for Woods-Saxon and harmonic oscillator density profiles along with $\alpha$-cluster structure considered in this study. These are shown in Figure~\ref{fig11}.  $v_2$ is found to increase at low-$p_{\rm T}$ and saturates at intermediate-$p_{\rm T}$. The qualitative trend of elliptic flow as a function of $p_{\rm T}$ in O+O collisions is found to be similar to that is observed by ALICE experiment in p+Pb collisions at $\sqrt{s_{\rm{NN}}}$ = 5.02 TeV~\cite{ALICE:2013snk}. Within uncertainties, $v_2$ is found to be similar for both Woods-Saxon and harmonic oscillator type density profiles for (0-5)\% and (20-30)\% centrality classes. This indicates that, although there is a significant dependence of the initial energy density on the density profiles, the collectivity (both radial and elliptic flow) is less affected by the modification in these two density profiles. However, when compared with results for $\alpha$-cluster structure in oxygen nucleus, elliptic flow at intermediate-$p_{\rm T}$ is found to have a centrality dependent trend as compared to the other two density profiles considered in this study. It is found that, for the most central case, elliptic flow for the $\alpha$-cluster nucleus is slightly less compared to the Woods-Saxon type nucleus. However, this trend seems to reverse as one moves to mid-central case, where $v_2$ is quantitatively more for the $\alpha$-cluster nucleus. To get a further insight into $\alpha$-cluster structure in oxygen nucleus and the similarities/differences with different density profiles, a detailed study on the centrality and transverse spherocity dependence of elliptic flow~\cite{Mallick:2021rsd} needs to be performed. It would be also interesting to compare the predictions of elliptic flow from AMPT with upcoming experimental results in O+O collisions.

\section{Summary}
\label{section4}

In summary, we report the predictions for global properties in O + O collisions at $\sqrt{s_{\rm{NN}}}$ = 7 TeV using a multi-phase transport model (AMPT). We report the mid-rapidity charged-particle multiplicity, mean transverse mass, Bjorken energy density, pseudorapidity distributions, squared speed of sound, $p_{\rm T}$ spectra, and the kinetic freeze-out parameters as a function of collision centrality. The results are shown for both harmonic oscillator and Woods-Saxon nuclear density profiles along with $\alpha$-clustered structure incorporated for oxygen nucleus. With the change of the density profile from Woods-Saxon to the harmonic oscillator and with the implementation of $\alpha$-clustered in the nucleus, a modification of average charged-particle multiplicity is seen, which is also reflected in the initial energy density as one naively expects. However, other global properties show less dependence on the density profile considered in this work. In general, the initial energy density produced in all collision centralities in O+O collisions at $\sqrt{s_{\rm{NN}}}$ = 7 TeV stays higher than the lattice QCD predicted
value for a deconfinement transition, making oxygen nuclei collisions a potential case of light-nuclei collisions at the LHC energies to create a state of QGP. In addition, a substantial radial flow and a comparable freeze-out temperature with that of Pb-Pb collisions are also observed in the present analysis. Although a significant dependence of the initial energy density on the density profiles is seen, the collectivity (both radial and elliptic flow) is less affected by the modification in the density profiles. When compared to the results from the $\alpha$-clustered structure incorporated in the oxygen nucleus with different density profiles, the initial energy density, and charged-particle multiplicity are found to be similar for the $\alpha$-clustered structure and harmonic oscillator density profile. However, the magnitude of elliptic flow at intermediate-$p_{\rm T}$ for $\alpha$-cluster and Woods-Saxon type nucleus is found to reverse the trends when studied as a function of different centrality classes. It would be interesting to confront these results with the experimental observations, when available, to reveal the density profile of the oxygen nucleus best suitable to describe ultra-relativistic nuclear collisions.
\newline
\section*{Appendix A}
\label{appendix:a}

\begin{figure}[ht]
\includegraphics[scale=0.4]{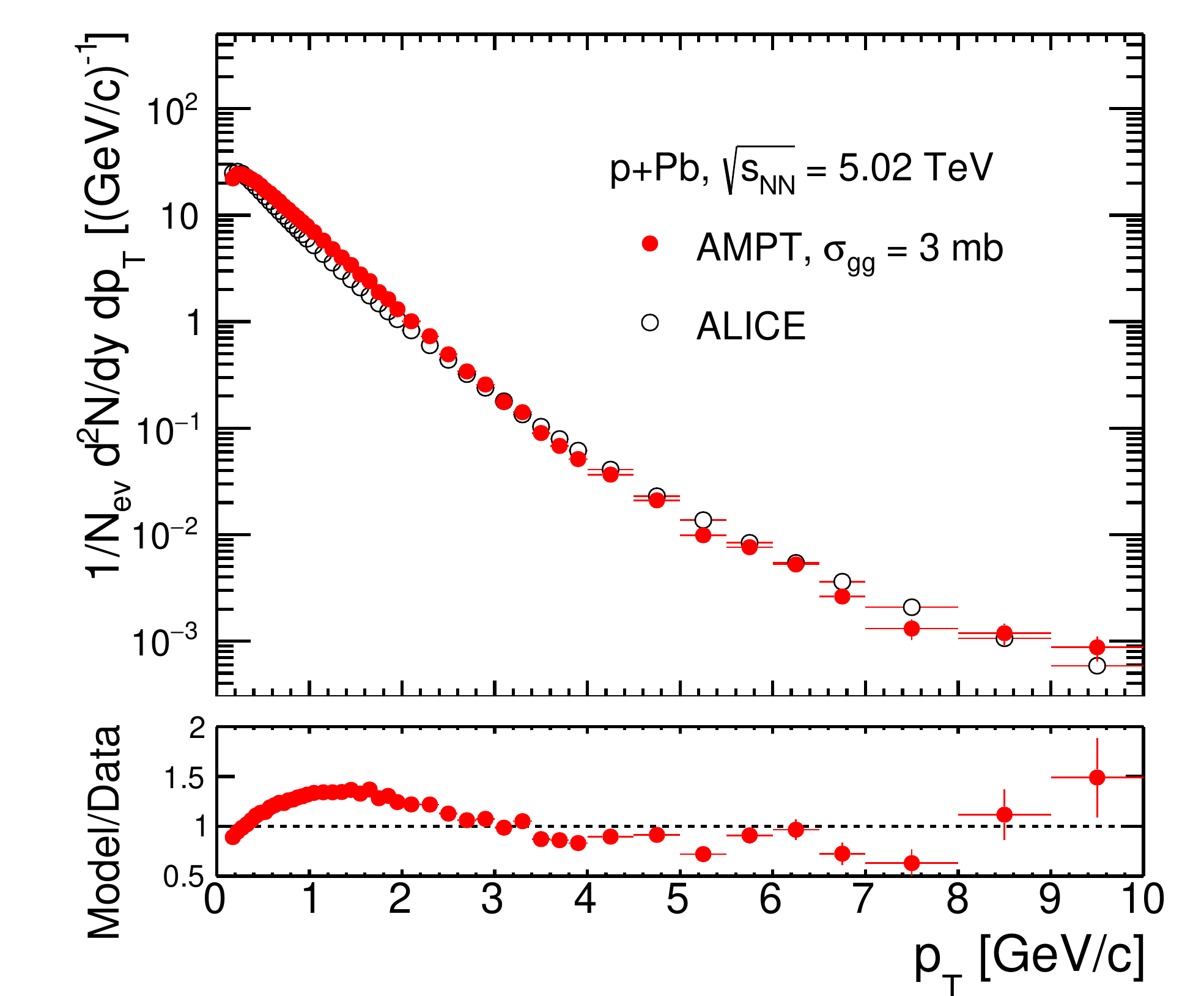}
\caption[]{(Color online) Comparison of ALICE data with the predictions from AMPT with $\sigma_{\rm{gg}}$ = 3 mb for p--Pb collisions at $\sqrt{s_{\rm{NN}}}$ = 5.02 TeV.}
\label{fig12}
\end{figure}

As the experimental data will be available for O+O collisions in the future (LHC RUN3), we have compared the charged particle $p_{\rm T}$-spectra from a collision system having a closer system size to O+O collisions $i.e.,$ minimum bias p+Pb collisions at $\sqrt{s_{\rm{NN}}}$ = 5.02 TeV with the predictions from AMPT. We have compared the AMPT predictions modifying the partonic scattering cross-section, $\sigma_{\rm{gg}}$ to different values such as 3, 5, and 10 mb. From this exercise, we found that fixing $\sigma_{\rm{gg}}$ = 3 mb, the AMPT predictions are closer to experimental data in p+Pb collisions at $\sqrt{s_{\rm{NN}}}$ = 5.02 TeV. Figure~\ref{fig12} shows the comparison of  ALICE data with the predictions from AMPT with $\sigma_{\rm{gg}}$ = 3 mb for p--Pb collisions at $\sqrt{s_{\rm{NN}}}$ = 5.02 TeV. It is found that the spectral shape from AMPT matches with the experimental data at high-$p_{\rm T}$ while at intermediate-$p_{\rm T}$, 10-30\% difference is observed. One should also note here that, to exactly match the experimental data, one can vary the tunes of the AMPT model, which is currently out of the scope of this manuscript. 

\section*{Appendix B}
\label{appendix:b}

\begin{figure}[ht]
\includegraphics[scale=0.4]{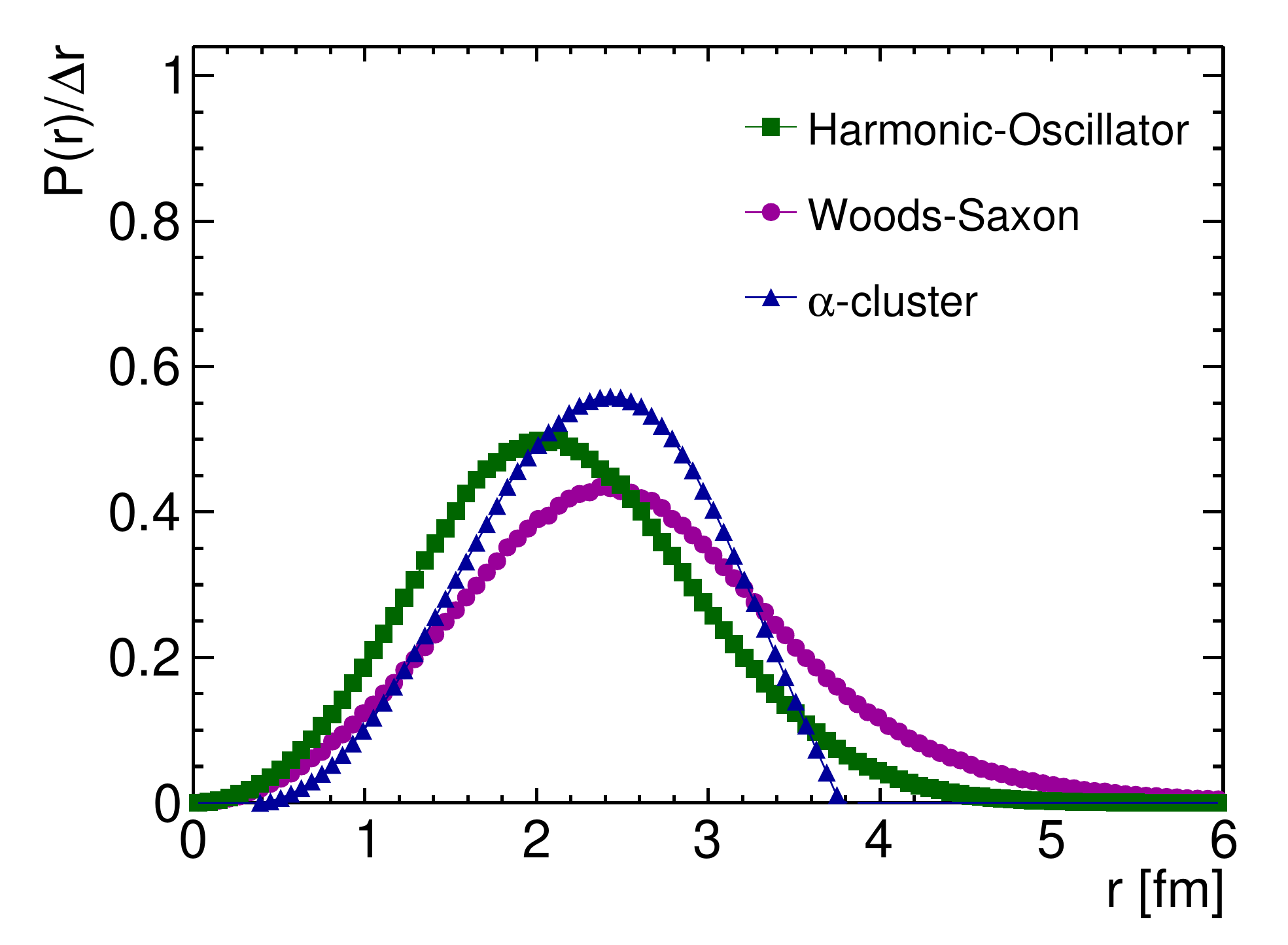}
\caption[]{(Color online) Probability of the radial position of the nucleons inside the oxygen nucleus. The rectangular (circular) markers represent the harmonic oscillator (Woods-Saxon) density profile, and the triangular marker represents the $\alpha$-clustered structure.}
\label{fig13}
\end{figure}

Figure~\ref{fig13} represents the probability of the radial positions of the nucleons distributed inside the oxygen nucleus for Woods-Saxon, harmonic oscillator density profiles, and $\alpha$-cluster structure. The $\alpha$-cluster structure seems to have a compact radial distribution of nucleons as compared to the other two density profiles. 


\section*{Acknowledgements}
We acknowledge Prof. Zi-Wei Lin for providing the necessary permission to modify the nuclear density profile in AMPT. DB acknowledges the financial support from CSIR, the Government of India. SP acknowledges the financial support from UGC, the Government of India. Raghunath Sahoo acknowledges the financial support under the CERN Scientific Associateship, CERN, Geneva, Switzerland, and the financial grants under DAE-BRNS Project No. 58/14/29/2019-BRNS of the Government of India. ST acknowledges the support under the INFN postdoctoral fellowship. A. N. M. thanks the Hungarian National Research, Development and Innovation Office (NKFIH) under the contract numbers OTKA K135515, K123815, and NKFIH 2019-2.1.11-TET-2019-00078, 2019-2.1.11-T ET-2019-00050 and the Wigner GPU Laboratory. The authors would like to acknowledge the usage of resources of the LHC grid computing facility at VECC, Kolkata, and the usage of resources of the LHC grid Tier-3 computing facility at IIT Indore.

%

 


\end{document}